\input epsf
\documentstyle[prb,eqsecnum,multicol,aps]{revtex}
\tighten
\draft
\begin{document}

\date{June 7, 1996}
%\date{\today}
\title{
Vortices in a thin film
 superconductor with a spherical geometry}
\author{M.\ J.\ W.\ Dodgson and M.\ A.\ Moore}
\address{Theoretical Physics Group \\
Department of Physics and Astronomy \\
The University of Manchester, M13 9PL, UK}
\maketitle
\draft 

\begin{abstract}
We report results from Monte Carlo simulations of a thin film superconductor 
in a spherical geometry within the lowest Landau level approximation. 
We observe the absence of a phase transition to a low temperature vortex solid 
phase with these boundary conditions; the system remains in the vortex liquid 
phase for all accessible temperatures. The correlation lengths are measured 
for phase coherence and density modulation. 
Both  lengths display identical temperature dependences,
 with an asymptotic scaling form consistent with  a 
continuous zero temperature transition. This contrasts with the first order 
freezing transition which is seen in the alternative quasi-periodic boundary 
conditions. 
The high temperature perturbation theory and the ground states of the
spherical system suggest that the thermodynamic limit of the spherical 
geometry is the same as that on the flat plane. We discuss the 
advantages and drawbacks of simulations with different geometries,
 and compare with current experimental conclusions. The effect of 
having a large scale
 inhomogeneity in the applied field is also considered.

\end{abstract}


\begin{multicols}{2}
\section{Introduction}

The existence and nature of melting transitions in two-dimensional (2D) 
systems 
remains a topic of interest and uncertainty in both theory and experiments. In
particular, the melting of the vortex lattice in a thin film superconductor has
received much attention.
A continuous melting 
transition from an 
Abrikosov lattice to a vortex liquid via a phase of hexatic order,
as in the Kosterlitz-Thouless-Halperin-Nelson-Young (KTHNY) theory,
was first proposed over fifteen years 
ago.\cite{fisher} A ``transition'' in the current-voltage response of 
thin film superconductors has been seen in several 
experiments.\cite{gammel,berghuis,yaz}
The vortex system becomes depinned above some transition temperature  
(the depinned vortices will dissipate energy as they move, leading to an 
Ohmic response) and the rise of the resistivity of the film 
is presumed to be caused by the  melting of the vortex lattice. However, it 
seems that this transition temperature varies with the strength of the pinning 
centers in the thin films. 
For films of amorphous NbO the pinning is  weak and no features can 
be detected in the current-voltage characteristics which may be attributed 
to a ``transition''. This has been interpreted as being due to 
 the absence of  the Abrikosov vortex lattice.\cite{nikulov}

The superconducting state arises from macroscopic correlations in the phase 
 of the Cooper pair wave function, or equivalently the 
Ginzburg-Landau order parameter. This phase coherence is usually a 
consequence of  off-diagonal long range order (ODLRO), i.e.\ non-vanishing 
$\langle\psi\rangle$, where $\psi$ is the superconducting order parameter. 
It has been predicted that thermal 
excitations of the shear modes of the vortex lattice should destroy ODLRO in 
the crystalline state and the length scale over which phase 
coherence persists has been estimated.\cite{moore,oneill}

O'Neill and Moore have performed Monte Carlo (MC) simulations of thin film 
superconductors within the lowest Landau level (LLL) 
approximation using a spherical geometry to 
minimize finite size boundary effects.\cite{oneill} 
No freezing transition out of the 
vortex liquid phase was found in the temperature region investigated. The 
correlation function associated with the positions of the vortices was 
calculated and the associated length scale of these lattice-like
correlations was found 
to diverge only in the zero temperature limit. It was noted that this 
happened on a similar  length scale 
over which there was expected to be phase coherence from the
arguments of Refs.~\onlinecite{moore}~and~\onlinecite{oneill}.

In the work of O'Neill and Moore, 
the superconducting film existed on the surface of a sphere,
and this geometry is the one employed here.
The
magnetic field perpendicular to the surface is generated by a magnetic 
monopole placed at 
the center of the sphere.\cite{dirac} 
The strength of the monopole must be quantized to have a well defined 
superconducting order parameter, 
but the strength of 
the field at the surface may be continuously varied by changing the 
radius of the sphere. 
The LLL approximation, 
which assumes an infinite magnetic screening length, 
works well in the 2D case where the effective screening length is 
$\Lambda_{eff} = 2\lambda^2/d$ because of the magnetic field outside the
superconductor.\cite{pearl} $\Lambda_{eff}$ becomes large as the
film thickness $d$ 
is reduced; $\lambda$ is the bulk magnetic penetration length.
It is often the case that $\Lambda_{eff}$ is greater than the film 
dimensions for small $d$.

More recently, strong evidence for a first order transition at a finite 
temperature in an alternative finite size approximation with quasi-periodic 
boundary conditions (QP) on an infinite plane has come from MC 
simulations.\cite{tesanovic,kato,humac,sasik,sst}  
This model has a periodic 
amplitude of the order parameter and a fixed change in the phase as a unit 
cell containing $N$ vortices is crossed (the change in phase depends on the 
choice of gauge). The quasi-periodicity in the phase 
places some constraints 
on the motion of the vortices, but it is not obvious that this will change 
the thermodynamic properties of the system. The observed transition 
temperature is within the range investigated on the sphere. This raises the 
question of whether or not there really is no transition on the sphere. 
If so, then what are the reasons that different 
boundary conditions give different results?

If there is a freezing transition, at least on the plane, does this 
contradict the theoretical prediction of no ODLRO at finite temperatures? 
Apparently not, as the work of Sasik, Stroud and Tesanovic\cite{sst} 
(SST) has concluded 
that in the thermodynamic limit even the low temperature state below the 
first order transition has no ODLRO, despite the presence of ``Bragg peaks'' 
in 
the correlations of the order parameter density (the peaks will not be 
delta-functions but the algebraic singularities expected for a 
2D crystalline system\cite{nelson}).
These conclusions of SST are taken as support for the ``charge-density wave''
phase (with a modulation in the Cooper pair density, but no ODLRO)
proposed by Tesanovic.\cite{tesan} 

In this work we again fail to find a phase transition in the 
spherical geometry. We have investigated the correlations associated 
with both the {\em phase} and the {\em amplitude} of the superconducting 
order parameter.
One advantage of the spherical geometry is that the liquid state may be 
investigated over a larger range of temperatures. This has allowed us to 
compare the growing length scales of the phase and amplitude correlations over 
a large variation in temperature.

Section~\ref{sec:method} contains the formulation of our model.
In section~\ref{sec:pert} it is shown how
the high temperature perturbation theory on the spherical geometry gives the
same free energy per vortex
 as the conventional theory in the thermodynamic limit.
In Section~\ref{sec:ground} 
we describe how the  ground state of the vortex 
system on the sphere is affected by the necessity of containing twelve
disclination defects within the triangular lattice. We have previously 
concluded that these defects with long range strains will give a finite 
energy cost
per vortex relative to the flat plane ground state even in the thermodynamic
limit.\cite{dodgson} However, the presence of
dislocation defects were not considered, and these can screen the strains
of the disclinations to recover the thermodynamic limit of the triangular
lattice on a flat plane. This screening is described in the framework
of elasticity theory, and the numerical evidence for 
dislocations in the ground state is reexamined.
In Section~\ref{sec:density} our calculations on the four-point 
density correlation function of the vortex liquid are shown. This is followed 
by our investigations on the phase coherence in Section~\ref{sec:phase}.

The spherical geometry is in principle realizable using the end
of a long thin 
solenoid to approximate the field from a monopole and placing it inside
a spherical superconducting film. However there would always 
be a problem in the placement of the solenoid which would lead to a large scale
inhomogeneity in the field strength at the spherical surface.
Additional simulations are  described 
in Section~\ref{sec:inhom}, where we add just such
 an inhomogeneous magnetic
field.

Finally, in the light of  different conclusions from simulations of
the LLL system in different geometries, we discuss some of the
problems in these geometries in Section~\ref{sec:discussion} 
and comment on the
experimental situation.
Some conclusions are drawn in Section~\ref{sec:conclusions}.

\section{Description of the Model}  \label{sec:method}

We model a thin film superconductor by a spherical shell of radius $R$ and
thickness $d \ll R$. An external magnetic field $\hbox{\bf H}(\hbox{\bf r})=
\hbox{\bf \^r}H(r)$,                       
always directed perpendicular
to the shell's surface is created by a magnetic monopole at the center of the
sphere. We neglect all fluctuations in the magnetic induction so that it always
has the mean value $B=\mu_0H(R)$ on the superconducting surface. 
Dirac's quantization condition\cite{dirac} (assuming that the basic unit of 
charge is that of a Cooper pair) requires that the total flux through the 
surface is an integer multiple of flux quanta $Nh/2e$.
Our choice of
gauge that satisfies $\nabla\times\hbox{\bf A}=\hbox{\bf B}$ is 
$\hbox{\bf A}\equiv
(A_r,A_\theta,A_\phi)=(0,0,BR\tan{\theta /2})$. We also assume that the
superconductor is  described by a Ginzburg-Landau (GL) complex order 
parameter $\psi$ with a phenomenological free energy Hamiltonian given by
\begin{equation} 
{\cal H}[\psi]=
\int d^3r \left[
\alpha(T) {|\psi|}^2                 
+ \frac{\beta}{2}{|\psi|}^4 
+ \frac{1}{2m} \psi^* {D}^2 \psi \right],\label{eq:glenergy}
\end{equation}  
where $D^2=\hbox{\bf D.D}$ and 
$\hbox{\bf D} = -i\hbar \hbox{\boldmath $\nabla$}
 -2e\hbox{\bf A}$ is the gauge invariant derivative operator. 
The LLL approximation consists in expanding this order
parameter in terms of the eigenstates of ${ D}^2$, and then only
keeping the degenerate level of states with the lowest eigenvalue. 
If we only
consider the two dimensions on the surface of the sphere then the subspace of
this LLL is spanned by $N+1$ orthonormal functions:\cite{roysing}
\begin{equation} 
\psi_m(\theta,\phi)=
h_m e^{im\phi}
\sin^m ({\theta}/{2})\,
\cos^{N-m} ({\theta}/{2})  ,
\end{equation}  
with $m=0,\ldots N$ and $h_m={[ (N+1)!/4\pi R^2m!(N-m)!]}^{1/2}$.
We write the order parameter as $\psi(\theta,\phi)= Q\sum
v_m \psi_m(\theta,\phi)$ with $Q=(\Phi_0 k_B T/\beta d B)^{1/4}$, and measure
lengths in units of the magnetic length $l_m=(\Phi_0/2\pi B)^{1/2}$ which 
fixes $R=(N/2)^{1/2}$. [Note that the lattice spacing $l_0$ in a triangular 
vortex lattice will be given by 
$l_0=(4\pi/\sqrt{3})^{1/2}l_m\simeq 2.69  l_m$.]
The LLL Hamiltonian can then be
written as
\begin{eqnarray}    \label{eq:oneill}
&&{\cal H}\left( \{u_m\} \right)=k_B T
\alpha_T^2\times\\
&&\left[ \hbox{sgn}{(\alpha_T)}\sum_{m=0}^N u_mu_m^*+
\sum_{p,q,r,s=0}^N w_{pqrs} u_pu_qu_r^*u_s^* \delta_{p+q,r+s} \right],
\nonumber
\end{eqnarray}
with $u_m=v_m/{|\alpha_T|}^{1/2}$.  
The quartic coupling term is given by\cite{oneill}
\begin{equation}
w_{pqrs}=\frac{(N+1)^2}{2N(2N+1)}f_{pq}f_{rs} ,
\end{equation} 
where ${f_{pq}}^2=C_p^NC_q^N/C_{p+q}^{2N}$ and $C_i^j=j!/[i!(j-i)!]$ 
is the binomial 
coefficient.

Our effective
reduced temperature variable is
\begin{equation}    
\alpha_T=\frac{dQ^2}{k_BT}\left[\alpha(T)+\frac{eB\hbar}{m}\right]
=\frac{dQ^2}{k_BT}\alpha'
\left[-1+\frac{T}{T_{c0}}+\frac{H}{H_{c2}(0)}\right].
\end{equation}
where $\alpha'=T_{c0}(d\alpha/dT)_{T=T_{c0}}$ and
$T_{c0}$ is the mean field transition temperature. $H_{c2}(0)$ is the 
straight line extrapolation to zero temperature of the upper critical field, 
within mean field theory.
For better computational efficiency, Eq.~(\ref{eq:oneill}) maybe rewritten
as\cite{hanlee}
\begin{equation}  \label{eq:ham}
{\cal H}\left( \{u_m\}\right) =k_B T
\alpha_T^2\left[\hbox{sgn}{(\alpha_T)} \sum_{m=0}^N |u_m|^2
+\frac{1}{2N}\sum_{p=0}^{2N} |U_p|^2\right]  ,
\end{equation}  
where $U_p=2\pi N\sum_{q=0}^N B^{1/2}(2N-p+1,p+1)h_qh_{p-q}\Theta(p-q)
\Theta(N+q-p)u_qu_{p-q}$ and $B(x,y)=\Gamma(x)\Gamma(y)/\Gamma(x+y)$ 
is the Beta function. $\Theta(q)$ is the Heaviside step function.

An alternative description of the order parameter on the sphere in the LLL 
approximation is possible if we write (following Ref.~\onlinecite{oneill})
$\psi(\theta,\phi)=Q\sum_{m=0}^N v_m\psi_m(\theta,\phi)=
Q\cos^N(\theta/2)\sum_{m=0}^Nv_mz^m$, 
where $z=\tan(\theta/2)e^{i\phi}$. The $N$th order polynomial may be written 
as a product of $N+1$ terms to give:
\begin{equation}\label{eq:prod}
\psi(\theta,\phi)=C\cos^N(\theta/2)\prod_{i=1}^N(z-z_i).
\end{equation}
Therefore the order parameter is determined by the positions of the $N$ zeros 
(vortices) $\{z_i\}$ along with an overall complex amplitude $C$. It is 
shown in Appendix~\ref{ap:inv} that the GL free energy Hamiltonian depends 
only on the {\em relative} positions of these vortices plus the overall 
complex amplitude, and not on the choice of coordinate system $(\theta,\phi)$ 
that arose from our choice of gauge. This translational invariance is 
assumed throughout this paper. Note that there are no constraints 
on the allowed positions $\{z_i\}$ which is not the case with  QP 
boundary conditions.

The properties of these systems are determined by the partition function
$Z=\int\prod_mdu_mdu_m^* \exp{[-{\cal H}(\{u_m\})/k_BT]}$. We investigate these
properties, 
in particular correlations in the  ``density'' $|\psi |^2$
and the phase $\arg{(\psi )}$, 
with the Metropolis algorithm which uses an MC method to
sample the phase space.\cite{metro} 
We have made measurements for runs up to $10^7$ MC steps for system sizes 
as large as
$N=400$. We have mainly concentrated on the temperature range between 
$\alpha_T=-2$ 
and $-12$. 
We have also used standard minimization
routines to investigate the ground states of this model.

\section{High temperature perturbation theory on the sphere}\label{sec:pert}

A perturbation expansion has been developed about the Gaussian limit
of (\ref{eq:glenergy}) 
at $\alpha_T\rightarrow +\infty$
to approximate various properties of 
the LLL system.\cite{ruggeri}
In two dimensions the geometry used has always been a flat plane.
The series may be represented by Feynman diagrams and many quantities 
in two and three dimensions have been evaluated to high order.\cite{HMM}
The series can extend to negative $\alpha_T$ as long as the propagator
is renormalized at least to the level of the Hartree-Fock approximation 
(HF).

We consider the perturbation series for the free energy using the 
spherical geometry that leads to the Hamiltonian (\ref{eq:ham}).
The usual treatment follows for the Feynman diagrams in a $\phi^4$ theory,
with directed
lines representing the gaussian propagator $g_0=\langle
v_pv_p^*\rangle_0$ connecting vertices representing the interaction
$\beta_{pqrs}=w_{pqrs} \delta_{p+q,r+s}$, 
and  internal lines are summed from $0$ to $N$.
All diagrams with loops are removed by renormalizing the propagator in a 
self-consistent fashion. This is the HF approximation, and the renormalized 
propagator, $\tilde{g}$, must satisfy
\begin{equation}
\tilde{g}=g_0(1-2\tilde{g}^2S^0),
\end{equation} 
where $S^0=\sum_{q=0}^N\beta_{pqpq}$. On the sphere we find:
\begin{equation}
S^0_{sphere}=\frac{N+1}{2N}=S^0_{plane}(1+\frac{1}{N}),
\end{equation}
so that in the large $N$ limit on the sphere, the HF propagator approaches
that on the flat plane.

Exactly the same diagrams appear in the free energy expansion for the 
spherical geometry as those for the plane.
The contribution of each diagram with
$n$ vertices
to the coefficient of the $n$th order term of the expansion is a product 
of a combinatorial factor and the integral/sum over the ``momenta'' along 
each line of the interactions at each vertex.
The only differences on the sphere are these sums.

\begin{figure}[tbp]
\narrowtext           
\epsfxsize= 8cm
\begin{center}
\leavevmode\epsfbox{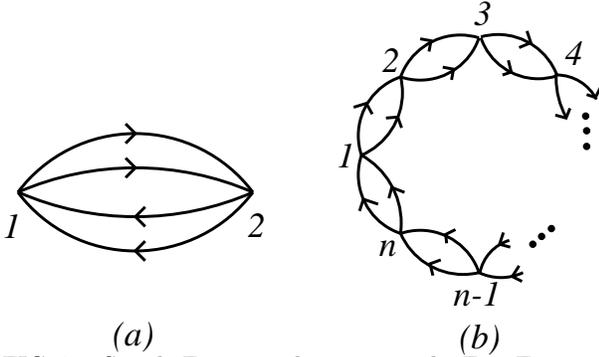}
\\
\caption{
Simple Feynman diagrams in the Free Energy series that we 
calculate on the spherical geometry: a) a second order diagram; 
b) part of the  $n$th order ring diagram.
\label{fig:feyn}}
\end{center}
\end{figure}

We now look at the lowest order Feynman graph without loops,
shown in Fig.~\ref{fig:feyn}a, which has 
two vertices connected by four lines. The sum that appears in this diagram
is $T^2=\sum_{pqrs=0}^{N}\beta_{pqrs}\beta_{rspq}$. On the plane this is 
equal to $T^2_{plane}=N/8$, and we find on the sphere:
\begin{equation}
T^2_{sphere}=\frac{(N+1)^4}{4N^2(2N+1)}=\frac{N}{8}[1+\frac{7}{2N}+
{\cal O}(N^{-2})],
\end{equation}
which also gives the result on the plane with a $1/N$ correction.

Generalizing our method to consider higher order terms, 
we calculate a particularly simple type
of diagram to arbitrary order: the ring diagrams shown in 
Fig.~\ref{fig:feyn}b.
For such a ring diagram of order $n$, the relevant sum is
$T^n=\sum_{p_i} \beta_{p_1p_2p_3p_4} \beta_{p_3p_4p_5p_6}\ldots
 \beta_{p_{2n-1}p_{2n}p_1p_2}$. We find on the sphere:
\begin{eqnarray}
T^n_{sphere}&=&\frac{(N+1)^{2n}}{2^nN^n(2N+1)^{n-1}}\nonumber\\
&=&T^n_{plane}[1+\frac{3n+1}{2N}+{\cal O}(N^{-2})],
\end{eqnarray}
where the result on a plane from
Ref.\onlinecite{mccauley}
is $T^n_{plane}=2^{1-2n}N$. So the contribution
to the free energy to $n$th order approaches that on the plane as long as
$n$ is much less than the system size $N$. 
Although we have not verified this for more complex diagrams, 
we expect a similar result. This is because the interactions at each vertex
have only $1/N$ (and higher power)
corrections on the sphere from the interactions on the
plane, so with $n$ vertices in a diagram, there should be a total correction
of order $n/N$.
Therefore as long as the 
perturbation theory is useful (ie. on the high temperature side 
of any phase transition) the thermodynamic limit on the sphere
should be the same as that for  the flat plane.
Intuitively, it is also reasonable that the free energy per vortex in 
the liquid phase should be the same on the flat plane and on the sphere. In the liquid phase, correlation lengths are finite (see Section~\ref{sec:sim}), 
and so long as they remain small compared to the sphere radius, then
 the liquid on the sphere is ``unaware'' of its curvature.
\section{Ground States on a Sphere}\label{sec:ground}

The ground state of the LLL vortex system on an infinite plane was long ago 
discovered  to
be the triangular lattice.\cite{kleiner} 
This is also the ground state on the finite systems used in the QP 
simulations when commensurate boundary conditions are chosen. 
However, on a sphere the closest configuration the vortices can make to an
ideal triangular lattice must contain twelve ``disclinations'', 
i.e.\ twelve vortices that
only have five nearest neighbors.

We have previously studied  the ground states by numerical minimization
of the Hamiltonian in Eq.~(\ref{eq:ham}) for different system 
sizes.\cite{dodgson} 
The following conclusions were made: 
there exist some values of $N$ (magic numbers)
 for which the ground state has a
 lower energy than nearby values, and the energies as $N\rightarrow\infty$
do not seem to converge to the infinite plane value. 
Extrapolation of our numerical
results gives
an extra energy on the sphere in this large $N$ limit of $0.2(4)$\% of 
the ground state energy per vortex on the infinite plane.
The vortex
configurations of these ground states make up a triangular network, but
with twelve five-coordinated centers. 
The result that as $N\rightarrow\infty$ the ground states do not 
look the same as for an infinite plane may be explained by the 
presence of the disclinations (five-coordinated
centers), which are defects with long range strains.\cite{seung} 
Therefore there will always be a
finite fraction of the vortex system that is distorted from an ideal triangular
lattice. 
This is supported by calculations within elasticity theory (which assumes only
small distortions) that give a similar finite energy cost of the distortions
on a sphere in the large $N$ limit when an icosahedral arrangement
of the disclinations is assumed.\cite{dodgson}

This result of an extra energy per vortex in the thermodynamic limit would 
be equally valid at finite temperatures as long as there is a non-zero 
shear modulus (ie. in a solid phase). If it were true there could be a 
paradox: 
Imagine a 2D system with a continuous phase transition
from a liquid to a crystalline solid phase. Above the transition the 
free energies per vortex have been shown in Section~\ref{sec:pert}
to be the same on the sphere and the plane and so this 
transition must remain at the same temperature
in each case. As at the transition 
the free energy is continuous, then the free energies in the 
solid state must also be equal.

In fact, one can do a better job on the ground states on the sphere by 
having a number of dislocations (which are 
bound pairs of five-fold and seven-fold disclinations) in addition to the 
twelve free disclinations. This means
 we have to
take away the restriction that every 
vortex is six-fold coordinated other than the twelve five-fold centers
used in the elasticity calculation of Ref.~\onlinecite{dodgson}.
There is no topological problem with this as one can pull an opposite 
pair of dislocations out of a perfect lattice.
 That extra dislocation defects  may lower the energy may be seen in two
different ways. First, the interaction energy between a dislocation and a
disclination may be negative, depending on the direction of the dislocation
and this interaction energy grows faster with system size than the self energy
of the dislocation. Alternatively, consider  a region on a sphere bounded by
three great circles that intersect at neighboring disclinations. Even in the
large $N$ limit this region will be distorted from a triangle (all three of 
its angles will be $2\pi/5$) and so a perfect triangular lattice in this region
will be incommensurate with the boundaries, where a 
series of dislocations will occur. So we have a picture of a ground state
in the large $N$ limit made up of twenty regions of perfect lattice, but with
sides containing dislocations, and corners containing disclinations.

There are still competing effects which discourage dislocations, and which 
may explain why they were not seen in Ref.~\onlinecite{dodgson}. First there
is the curvature of the sphere which favors the distortions of a 
disclination, but which decreases as $1/R^2$. Second, there is the positive 
self energy of a dislocation, which is generally expected to increase as the
logarithm of the system size.\cite{nabarro} 
Third there is the  interaction between
two
dislocations which is directional, and approximately 
proportional to the logarithm of 
the distance between them. To estimate the density of dislocations needed
to reduce the strain energy of the disclinations we use the framework of 
elasticity theory.

\subsection{Elasticity calculation of defects on a sphere}

The elasticity theory of twelve disclinations on a sphere is described in 
Ref.~\onlinecite{dodgson}. The method is briefly explained here and then 
generalized to allow the inclusion of dislocations. We consider small 
strains from an ideal lattice, $u_{ij}(\hbox{\bf r})$. These strains are
 related to a stress field by Hooke's law, $\sigma_{ij}=2\mu u_{ij}
+\lambda u_{kk}\delta_{ij}$. For the LLL system the elastic 
constant $\lambda$ diverges (the system is incompressible)
and the shear modulus $\mu$ is given by 
$\mu=0.48k_BT\alpha_T^2/4\pi\beta_A^2l_m^2$. The stress tensor has zero 
divergence which allows the problem to be expressed in terms of the 
Airy stress function, $\chi$, defined by 
$\sigma_{ij}=\epsilon_{ik}\epsilon_{jl}\partial_k
\partial_l\chi$. Five-fold disclinations are defined by the change in 
bond angle by $ 2\pi/6$ when a path encircles one such defect. Dislocations,
on the other hand, are defined by their Burgers vector, but they are also 
equivalent to a disclination dipole pair.
The stress field corresponding to a particular configuration of these 
topological defects is given by the solutions to the biharmonic equation
on a sphere:
\begin{equation}
\frac{1}{K_0}\nabla^4\chi=-\frac{1}{R^2} +\sum_{s=1}^{12}\frac{2\pi}{6}
\delta(\hbox{\bf r}-
\hbox{\bf r}_s) +\sum_d \epsilon_{ij}{b_i}^{d}\partial_j
\delta(\hbox{\bf r}-
\hbox{\bf r}_d),\label{eq:biharm}
\end{equation}
where $K_0$ is the 2D Young's modulus which in the LLL system is given by
$K_0=4\mu$. The free disclinations  are labeled by $s$,
and $d$ labels the dislocations with
Burgers vectors {\bf b}$^{d}$. In the large $R$ limit where we neglect
the energy of bending the lattice over the spherical curvature, the 
energy of a given configuration is
\begin{equation}
F_{el}=\frac{1}{2K_0}\int d^2r \,(\nabla^2\chi)^2
=\frac{1}{2K_0}\int d^2r\, {\sigma_{kk}}^2,
\end{equation}
where the integrals are over the surface of the sphere.

We will need to use coordinate frames rotated so that the
reference axis passes through the defect. For instance,
if a disclination is at the position $(\theta_s,\phi_s)$ in our original 
coordinate frame, a general point $(\theta,\phi)$ is described in the 
rotated frame
by the coordinates $\theta'(s)$ (the polar angle from the disclination)
and $\phi'(s)$ (the azimuthal angle in the rotated frame relative 
to the $\phi=\phi_s$ direction), where
\begin{eqnarray}
\cos\theta'(s)&=&\cos\theta\cos\theta_s +
\sin\theta\sin\theta_s\cos(\phi-\phi_s)\label{eq:theta'}\\
\sin\theta'(s)\sin\phi'(s)&=&\sin\theta\sin(\phi-\phi_s)\\
\sin\theta'(s)\cos\phi'(s)&=&\sin\theta\cos\theta_s\cos(\phi-\phi_s)
-\cos\theta\sin\theta_s.
\end{eqnarray}
For the reference direction of the azimuthal angle $\phi'(s)$ 
in the new frame also
to be rotated, a further transformation $\phi'(s)\rightarrow\phi''(s)$
is required (as in the standard 
treatment of Euler angles\cite{gold}). We use this
extra rotation for the dislocation 
defects, but for simplicity we do not define the transformation here.

The solution of Eq.~(\ref{eq:biharm}) 
in the absence of dislocations was found to be 
$\chi=\sum_{s=1}^{12} \chi_s$ where the $s$th disclination is at 
$(\theta_s,\phi_s)$ and
\begin{equation}
\nabla^2\chi_s(\theta,\phi)=\sigma_{kk}^{s}(\theta,\phi)
=\frac{K_0}{12}\{\ln\frac{1}{2}
[1-\cos{\theta'(s)}] +1 \}.
\end{equation}
The elastic energy cost of the twelve disclinations is $E_{12}=R^2
\int d\Omega [\sum_s \sigma_{kk}^{s}(\theta,\phi)]^2$, which is 
proportional to the area of the system, and gives a finite energy 
cost per vortex.

If we now consider additional dislocation defects, we may write
$\chi=\sum_s \chi_s+\sum_d\chi_d$ where
\begin{equation}
\frac{1}{K_0}
\nabla^4\chi_d=\epsilon_{ij}b_i^{d}\partial_j\delta(\hbox{\bf r}-
\hbox{\bf r}_d).
\end{equation}
By differentiating the solution for a disclination, we find:
\begin{equation}
\sigma_{kk}^{d}(\theta,\phi)=
\frac{K_0l_0}{4\pi R}\frac{\sin{\theta'(d)}
\sin{\phi''(d)}}{1-\cos{\theta'(d)}},
\end{equation}
where we take the size of the Burgers vector to be the lattice spacing 
$l_0$, and $\phi''(d)$ is the azimuthal angle about the axis through the 
dislocation measured from the direction of the Burgers vector.

Within elasticity theory, we can consider separately the contributions of 
the self energy of each defect, and the pairwise interaction energy 
between the defects. The self energy of a dislocation on a sphere is
\begin{equation}
E_d=
%\frac{1}{2K_0}\int d^2r\,(\sigma_{kk}^d)^2=
-\frac{K_0l_0^2}{32\pi}\left\{
1+\cos(a/R)+2\ln\frac{1}{2}[1-\cos(a/R)]\right\} +E_{core},
\end{equation}
where $a$ is a cutoff of the order of the lattice spacing (within $a$ of 
the dislocation, the linear elasticity theory must break down completely)
and $E_{core}$ is the energy of the distortion inside this cutoff. 
As expected, this becomes proportional to $\ln{(R/a)}$ in the limit of 
large $R$, where:
\begin{equation}
\lim_{R\rightarrow\infty}(E_d)=\frac{K_0l_0^2}{16\pi}[
2\ln(R/a) +2\ln{2}-1] +E_{core}.
\end{equation}

In the presence of more than one dislocation the total elastic energy will be
\begin{eqnarray}
E_{tot}&=& \frac{1}{2K_0}\int d^2r\,(
\sum_{s=1}^{12}\sigma_{kk}^s+
\sum_{d=1}^{N_d}\sigma_{kk}^d)^2\\
&=& E_{12}+N_dE_d+
\sum_{s=1}^{12}\sum_{d=1}^{N_d}E_{sd}(\hbox{\bf r}_s,\hbox{\bf r}_d,
\hbox{\bf b}^{d})\nonumber\\
&&+
\sum_{d=1}^{N_d}\sum_{d'>d}^{N_d}E_{dd}(\hbox{\bf r}_d,\hbox{\bf r}_{d'},
\hbox{\bf b}^{d},\hbox{\bf b}^{d'}),
\end{eqnarray}
which is a sum of the self energies plus pairwise interactions between 
every defect (the interactions between disclinations are included in $E_{12}$).

If we consider a disclination at $(\theta_s,\phi_s)=(0,0)$ and a 
dislocation at 
$(\theta_d,\phi_d)$
with a Burgers vector pointing perpendicular to the $\phi=\phi_d$ line 
(such that the five-seven pair in the dislocation point towards the 
disclination),
we find an interaction energy:
\begin{equation}
E_{sd}(\hbox{\bf r}_s,\hbox{\bf r}_d,
\hbox{\bf b}^{d})=
-\frac{K_0l_0R}{12}\left[
\frac{(1-\cos{\theta_d})}{\sin{\theta_d}}
\ln\frac{1}{2}[1-\cos(\theta_d)]\right].
\end{equation}
Allowing the dislocation to rotate by an angle $\gamma$ will give the above 
energy multiplied by a factor $\cos{\gamma}$. 
\begin{figure}[htbp]           
\epsfxsize= 8cm
\begin{center}
\leavevmode\epsfbox{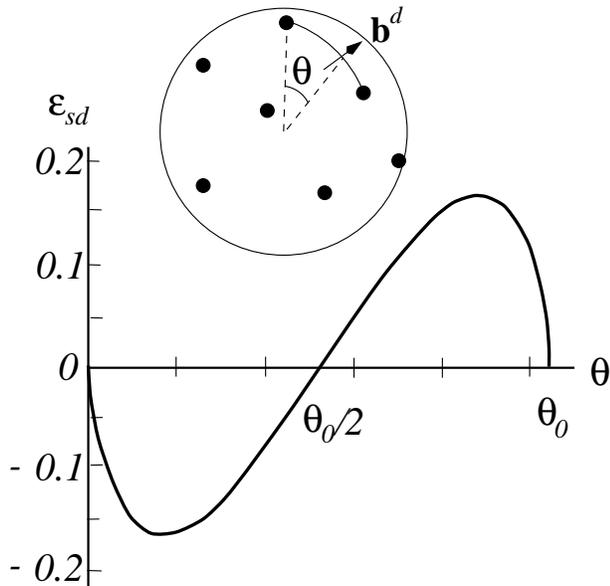}
\\
\caption{  The interaction energy $E_{sd}^{icos}=(K_0l_0R/12)\epsilon_{sd}$
between a dislocation and twelve 
disclinations in an
icosahedral configuration, 
where the dislocation is on a line between two 
neighboring disclinations, 
with a Burgers vector perpendicular to the line.
The polar angle $\theta_0$ between two neighboring disclinations 
is defined in the text.
The polar angle between the 
dislocation and the disclination is $\theta$ as shown in the diagram, 
where the spots represent the positions of disclinations on
the sphere.
  \label{fig:disloc}}
\end{center}
\end{figure}

The interaction energy between two dislocations, one at 
$(\theta_{d'},\phi_{d'})=(0,0)$, the other at $(\theta_d,\phi_d)$, with 
opposite Burgers vectors perpendicular to the line joining the dislocations, 
is found to be
\begin{equation}
E_{dd}(\hbox{\bf r}_{d'},\hbox{\bf r}_d,
\hbox{\bf b}^{d'},\hbox{\bf b}^{d})=
\frac{K_0l_0^2}{4\pi}\left[
1+\frac{\ln\frac{1}{2}[1-\cos\theta_d]}
{(1+\cos{\theta_d})}
\right].
\end{equation}
Again as expected, the total energy of a dislocation pair, $2E_d+E_{dd}$, 
in the large $R$ limit becomes proportional to the logarithm of the 
distance between the two opposite dislocations.

Notice that the interaction energy between a dislocation and disclination 
increases more rapidly with $R$  than the self energy and the 
dislocation-dislocation interaction energy. As $E_{sd}$ may be negative 
(depending on the orientation of the Burgers vector), there is an 
instability to the formation of dislocations, for large enough system size.
If we consider the twelve disclinations to be fixed at the corners of an 
icosahedron, we find the energy landscape of a single dislocation to have 
minima along the lines that join the disclinations. 
The energy $E_{sd}^{icos}$ (which is a sum of the interactions
between the dislocation and the twelve disclinations)
is shown in Fig.~\ref{fig:disloc}
as a function of position on one of these lines, 
with a Burgers vector that is perpendicular to the line.
The position is denoted by the polar angle $\theta$ 
from one disclination,
with a neighboring disclination located at 
$\theta=\theta_0=2\tan^{-1}[(\sqrt{5}-1)/2]$.
By symmetry, the energy is zero exactly on a disclination and at the 
midpoint between two disclinations, and there is one negative stationary point.

For an estimate of when dislocations should first start to appear in the 
ground states as $R$ increases, we consider the energy to put two 
dislocations of opposite Burgers vectors on each line joining disclinations, 
at the positions that minimize $E_{sd}$.
Such a configuration has an elastic energy cost of
\begin{eqnarray}
E_{tot}=&&E_{12}+K_0l_0^2[
-0.861(R/l_0)+2.39\ln(R/l_0) +0.907]
\nonumber\\&& +60E_{core},
\end{eqnarray}
where we assume $a=l_0$ and have only taken into account the 
dislocation-dislocation interactions between pairs on the same line.
If we neglect $E_{core}$ we find that $E_{tot}=E_{12}$ when $R=6.05l_0$ which 
corresponds to
$N\simeq 530$. However, with an estimate of $E_{core}\sim
K_0l_0^2/16\pi$, this critical 
value increases to $N\simeq 1000$. The critical value of $N$ is 
sensitive to the estimated values of the core energy, and the
neglect of extra dislocation pair interactions. 
Also, we have neglected the local curvature, which for small systems will
favor the presence of the disclination strains, and therefore disfavor 
dislocations that screen these strains.
Our estimates are compared
with numerical results later in this section.

Having established that dislocations may lower the total elastic energy 
cost on the sphere for large enough systems, it remains to show that large 
numbers of dislocations may screen out the strain field of the disclinations 
so as to cancel the $R^2$ dependence of the elastic energy. 
\begin{figure}[htbp]           
\epsfxsize= 8cm
\begin{center}
\leavevmode\epsfbox{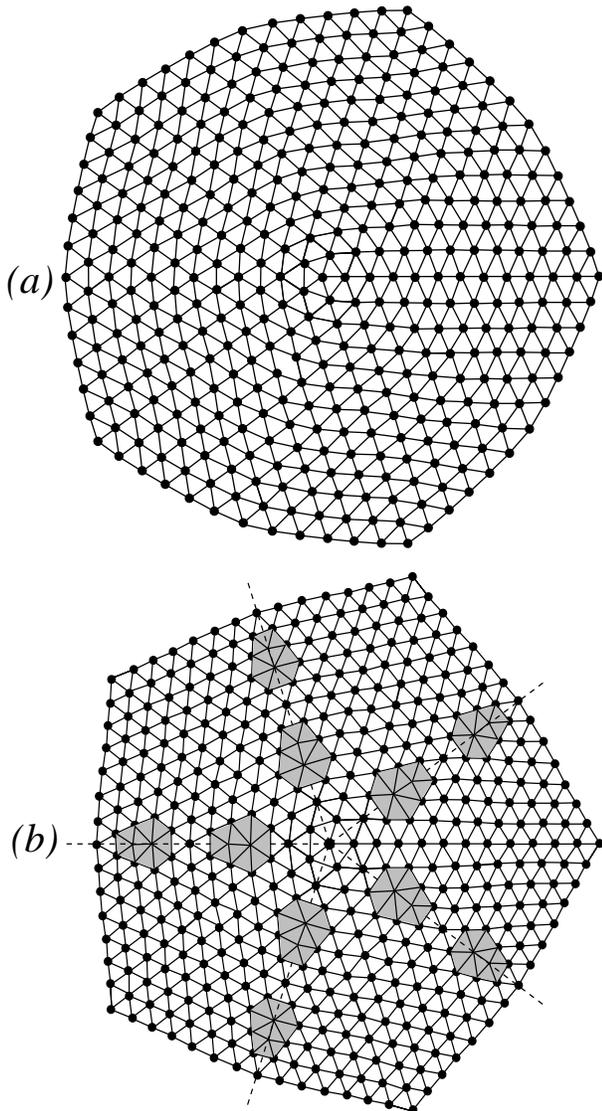}
\\
\caption{ (a) A disclination defect on a flat plane without dislocations.
(b) An example of dislocation series screening a disclination on a
flat plane.
 \label{fig:screen}}
\end{center}
\end{figure} 
We consider series of dislocations along the lines between 
the disclinations.  The sum of the strains from each defect may be 
approximated at large distances by an integral.
For such a series with equally spaced dislocations a distance $cl_0$ apart,
the resulting strain fields are those from a positive and a negative 
disclination at each end of the line, but with a charge of
size $q_s=1/c$ (where $q_s=+1$ for a five-fold disclination).
This charge arises from the ratio of the distance between the five-seven pair
in a dislocation to the distance between dislocations in the series.  
This is seen by writing the sum of the dislocation strain fields
as an integral. A series of dislocations running from $(0,0)$
to some point $(\Theta,0)$ will have a total strain field of
\begin{eqnarray}
\sigma^c(\theta,\phi)&\simeq&\frac{R}{cl_0}\int_0^{\Theta} 
\sigma^d_{\theta'}(
\theta,\phi)d\theta'\nonumber\\
&=&\frac{1}{c}\sigma^s_{\Theta}(\theta,\phi)-\frac{1}{c}
\sigma^s_{0}(\theta,\phi),
\end{eqnarray}
because the strain field $\sigma^d_{\theta'}(\theta,\phi)$ of a dislocation 
at $(\theta',0)$ with Burgers 
vector perpendicular to the $\phi=0$ direction is just the differential
of the strain field $\sigma^s_{\theta'}(\theta,\phi)$
of a disclination: $\sigma^d_{\theta'}=
l_0[\partial\sigma^s_{\theta'}/\partial(R\theta')]$.

If we imagine such series with $c=5$ between each disclination and the 
midpoints to neighboring disclinations, as is shown near one disclination
in Fig.~\ref{fig:screen}, 
then the strain field of each 
disclination will be exactly screened away. However, 
there will be an effective 
disclination of
charge $q_s=2/5$ at each midpoint
[the midpoint makes a polar angle $\theta_m=\theta_0/2$
with the disclination, so that $\tan{\theta_m}=(\sqrt{5}-1)/2$]. 
What is needed is a variable 
spacing 
that starts from $5 l_0$ at the disclination, but diverges at the midpoint.
We may require that the spacing 
has the scaling form $c(\theta')=5+(R/l_0)g(\theta')$ with the condition that
$g(0)=0$ (this form is chosen as any higher power of $R$ would lead to 
a spacing that increases faster than the length of the series).

In the large $N$ limit an exact cancellation of 
the $R^2$ terms in the energy is obtained.
The strain field from one series starting at $\theta=0$
and lying along $\phi=0$ is given by
\begin{eqnarray}
\sigma^c(\theta,\phi)&=&\int_0^{\theta_m} 
\sigma^d_{\theta'}(
\theta,\phi)c^{-1}(\theta')Rd\theta'\nonumber\\
&=&
\left[ \sigma^s_{\theta'}(\theta,\phi)c^{-1}(\theta')\right]^{\theta_m}_0
-
\frac{l_0}{R}\int_0^{\theta_m} 
 \sigma^s_{\theta'}(
\theta,\phi)
\frac{dg^{-1}}{d\theta'}d\theta'\nonumber\\
&=& -\frac{1}{5}\sigma^s_0 +\frac{1}{[5+(R/l_0)g(\theta_m)]}
\sigma^s_{\theta_m}
-\frac{l_0}{R}I(\theta,\phi),
\end{eqnarray}
where $I(\theta,\phi)=\int_0^{\theta_m}(dg^{-1}/d\theta')
\sigma^s_{\theta'}d\theta'$, which remains
 of order one as the system size increases.
In the large $R$ limit we write the strain field of the $i$th series 
that is connected to the disclination $j$ as $\sigma^c_{i,j}(\theta,\phi)
=-(1/5)\sigma^s_j(\theta,\phi) +g^{-1}(\theta_m)(l_0/R)
\sigma^s_{i,j,\theta_m}(\theta,\phi)
-(l_0/R)I_{i,j}(\theta,\phi)\equiv -(1/5)\sigma^s_j+(l_0/R)f_{i,j}$
where $i$ runs from one to five.
The elastic energy cost due to all the defects,
$E_{tot}=E_{12}+E_{sd}+E_{dd}$ is
\begin{eqnarray}
E_{12}&=&\frac{R^2}{2K_0}
\int d\Omega\,\sum_{j,j'=1}^{12}\sigma^s_j(\theta,\phi)
\sigma^s_{j'}(\theta,\phi)\nonumber\\
&\equiv& 
\frac{1}{2K_0}\sum_{j,j'=1}^{12}R^2\epsilon_{jj'}\nonumber\\
E_{sd}&=&-\frac{1}{K_0}\left(
\sum_{j,j'=1}^{12}R^2\epsilon_{jj'}-
l_0R\sum_{j,j'=1}^{12}\sum_{i=1}^{5}\int d\Omega\,\sigma^s_{j'}f_{i,j}\right)
\nonumber\\
E_{dd}&=&\frac{1}{2K_0}\sum_{j,j'=1}^{12}\Biggl( 
R^2\epsilon_{jj'}+
2l_0R\sum_{i=1}^{5}\int d\Omega\,\sigma^s_{j'}f_{i,j}\nonumber\\
&&\hspace{3.5cm} +2l_0^2
\sum_{i,i'=1}^{5}\int d\Omega\,f_{i,j}f_{i',j'}
\Biggr)\nonumber\\
E_{tot}&=&\frac{l_0^2}{K_0}
\sum_{j,j'=1}^{12}\sum_{i,i'=1}^{5}\int d\Omega\,f_{i,j}(\theta,\phi)
f_{i',j'}(\theta,\phi),
\end{eqnarray} 
which shows that all of the terms of order $R^2$ cancel, leaving terms of order
one which depend on the gradients of $g^{-1}(\theta')$
and the value of $g^{-1}(\theta_m)$. 
Our approximation in simplifying the effective charge at the midpoint to 
be proportional to $1/R$ ignores a correction of order $1/R^2$ to this charge.
As the energy of a disclination is 
proportional to the area of the system, the energy cost 
of this ignored extra charge is also of order one.
Therefore, the production of these
series of dislocations
of variable spacing, along the lines between disclinations, is an effective 
way of screening the diverging strains of the disclinations. The 
ground state energy per vortex in the large $N$ limit is the same as that
for an infinite flat plane apart from a correction proportional to
$N^{-1/2}$ due to the core energies of the dislocations.

\subsection{Numerical evidence for dislocations in ground states}

We now consider the numerical evidence for the above conclusions. It is sufficient to minimize the Hamiltonian of
Eq.~(\ref{eq:ham}) by minimizing the Abrikosov ratio
\begin{equation}
\beta_A=\frac{\langle|\psi|^4\rangle}{\langle|\psi|^2\rangle^2}
=\frac{\sum_{p=0}^{2N}|U_p|^2}{\left(\sum_{q=0}^N|u_q|^2\right)^2}.
\end{equation}
The triangular lattice on the flat plane has the value\cite{kleiner} 
$\beta_{A,0}\simeq 1.1596$.
In Ref.~\onlinecite{dodgson},
the ground states of system sizes up to $N=652$ 
were found numerically, and it was
suggested that there was a finite 
energy cost per vortex in the large $N$ limit, consistent with the
presence of twelve disclinations. The existence of magic numbers was observed 
and explained, although for large systems this only has a small effect on 
the total energy.

We have extended this investigation to larger systems;
this is 
motivated by the above elasticity results which suggest 
that above a critical size, dislocations will appear in the 
ground states and these will screen the strains of the disclinations to
recover the large $N$ limit ground state energy of the flat plane.
In Fig.~\ref{fig:screen} the sort of configuration is shown that we
expect to find near each disclination. This figure demonstrates just how 
large the spherical system must be to accomodate the dislocations and 
screen out the strains over a large region.

\begin{figure}[htbp]           
\epsfxsize= 8cm
\begin{center}
\leavevmode\epsfbox{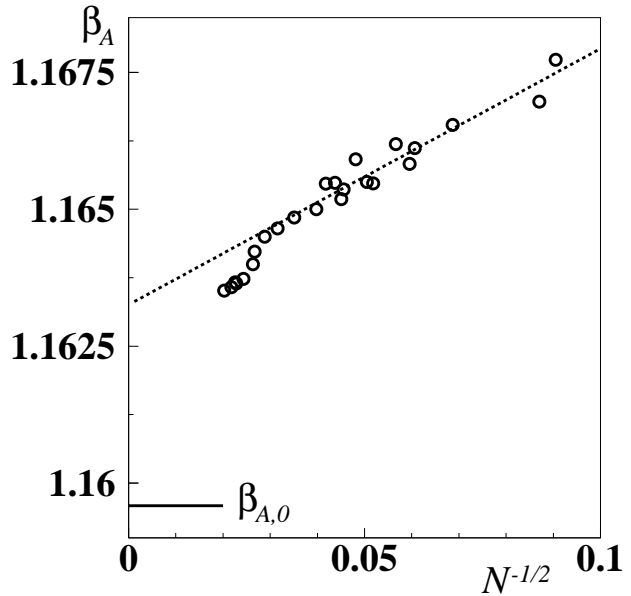}
\\
\caption{ The Abrikosov ratio of numerically found ground states for
different system sizes. Only magic number systems are shown. The straight line
fit to $N^{-1/2}$ is only on data for $N<1400$. Above this value, 
dislocations are seen in the ground states 
and $\beta_A$ falls under the fit.
 \label{fig:beta}}
\end{center}
\end{figure}

In Fig.~\ref{fig:beta} our numerical results are shown for 
ground states of systems larger than $N=652$ up to $N=2432$. We also 
include results from Ref.~\onlinecite{dodgson} for smaller system sizes. 
It is important to note that, as the systems get large, there is an 
increasing number of metastable states that the minimization routines 
may fall into. In each case we have taken the lowest energy after several
hundred runs, but this is always just an upper bound on the true ground state 
energy. 

Plotting 
the Abrikosov ratio
against $N^{-1/2}$ the results for $N<1200$ fall on a straight line that 
extrapolates to give a finite difference for the flat plane as 
$N\rightarrow\infty$. However, this trend changes above $N=1200$, with 
energies falling below the straight line.
The results are explained by the reasoning that once systems are large
enough for dislocations to overcome their self energy and mutual
 repulsion, they may screen the strains of the disclinations and reduce 
the total ground state energy. 
In the numerical ground states with $N>1200$ we observe dislocations
near the disclinations with the correct orientation of Burgers vector
(the distribution of the dislocations may be irregular due to the
difficulties in locating the true global ground state).

The elasticity calculation combined with 
this numerical work is strong evidence that the ground state energy per 
vortex in the thermodynamic limit is the same on the sphere as on the 
flat plane. 
It should be mentioned however that
  for the system sizes we look at in MC simulations,
the
ground states have a higher energy per vortex by $0.5$-$1$\% of the ground 
state energy on an infinite plane.
The jump in the energy at the transition observed in the QP systems is 
of a similar order of magnitude ($\sim 1\%$) as the energy difference of the 
spherical ground states from the QP ground states.
It must be a possibility that the raising of the ground state energy inhibits 
a freezing transition to the low temperature state, which could explain 
the apparent absence of a phase transition in our simulations on the sphere.

\section{Simulations on the sphere}\label{sec:sim}

We now go on to measure the 
correlations in the 2D vortex system using MC simulations.
In the following, the system sizes we often use correspond to the magic numbers
with particularly low energy ground states. This is because a phase transition
would be most likely where the ground state is least frustrated. Also, where 
we start the simulations in the ground state, we have most confidence that
our numerical ground states are the correct ones in the magic number cases.
However, we fail to observe commensurability effects in the MC simulations.
The magic numbers are a property of the ground state, and do not affect
the vortex liquid over the temperature range investigated.

\subsection{The Density Correlation Function}\label{sec:density}

\begin{figure}[htbp]           
\epsfxsize= 8cm
\begin{center}
\leavevmode\epsfbox{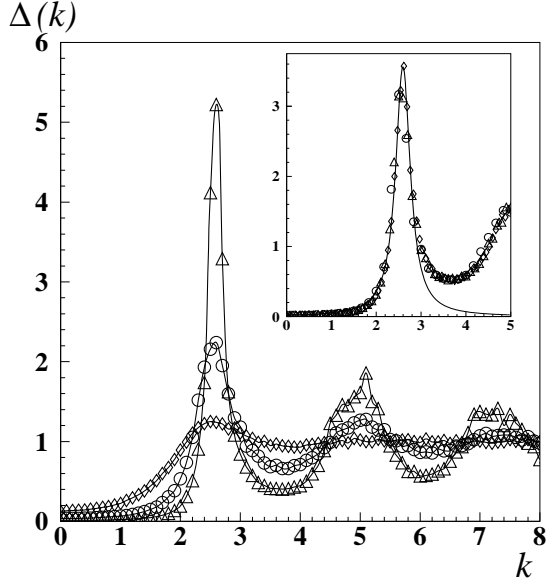}
\\
\caption{ Density correlation function with $N=200$ for reduced temperatures
\hbox{$\alpha_T=-2,-5$ and $-10$} (the peak heights increase with 
lower temperatures).
The inset shows the density correlation function at $\alpha_T=-8$ with 
\hbox{$N=72,200$ and
$400$}. The solid line is a Lorentzian fit when $N=400$.
 \label{fig:1}}
\end{center}
\end{figure}

O'Neill and Moore have investigated
the correlation function associated with vortex positions.\cite{oneill} 
This method is  unwieldy as it requires finding the
roots of an $N$th order polynomial at every measurement.
It is more convenient to look at the four point density correlation function
(briefly investigated by Lee and Moore\cite{hanlee}
and related to the correlator studied on the QP systems\cite{humac})
\begin{equation}
\chi (\hbox{\bf r}-\hbox{\bf r}')=
\langle  {|\psi(\hbox{\bf r})|}^2
{|\psi(\hbox{\bf r}')|}^2   \rangle
- \langle  {|\psi(\hbox{\bf r})|}^2 \rangle 
\langle{|\psi(\hbox{\bf r}')|}^2 \rangle  .
\end{equation}
This is most revealing in reciprocal space. On a spherical surface this means
that we need to decompose the function into spherical harmonics:
\begin{equation}
\chi_l^m =\int d\Omega d\Omega' 
\chi (\hbox{\bf r}-\hbox{\bf r}')
{Y_l^m}^*(\theta,\phi)
Y_l^m(\theta',\phi') .
\end{equation}
Within the LLL subspace, this correlator can be written as thermal averages
over the coefficients $u_p$:
\begin{eqnarray} \label{eq:recip}
\chi_l^m =\frac{\pi N^2{\overline{\rho}}^2}{{\left(   \sum_{p=0}^N
 \langle u_p^* u_p  \rangle \right)}^2}
\Biggl[
\sum_{p,q,r,s=0}^N&&
I_{p,q,l}^m I_{r,s,l}^m \langle u_p^* u_q u_r^* u_s \rangle \nonumber\\
&& \hspace{-5mm}
- \frac{\delta_{l,0}}{\pi N^2}
{\left(   \sum_{p=0}^N
 \langle u_p^* u_p  \rangle \right)}^2 \Biggr],
\end{eqnarray}
with $I_{p,q,l}^m$ given in Appendix~\ref{ap:struc} and $\overline{\rho}$ 
equal to 
the mean density $\overline{\rho} =\langle  {|\psi|}^2 \rangle$. 
To make comparison with
different temperatures we look at the function 
$C_l^m=\chi_l^m/{\overline{\rho}}^2$
and then normalize this function by its infinite temperature limit
$C_{l,\,\infty}$ (also given in Appendix~\ref{ap:struc}). 
We average our measurements over $m$, 
so our investigations center on the function
$\Delta(k)=\sum_{m=-l}^l C_l^m /[(2l+1) C_{l,\,\infty}]$
with $k=l/R$. 
$\Delta (k)$ is the same correlator that has been looked at in a plane 
geometry by 
Hu and MacDonald\cite{humac} as well as by Yeo and Moore,\cite{yeo} 
but modified for a sphere.
Examples of this function for different $\alpha_T$ and $N$ are
shown in Fig.~\ref{fig:1}.

\begin{figure}[htbp]           
\epsfxsize= 8cm
\begin{center}
\leavevmode\epsfbox{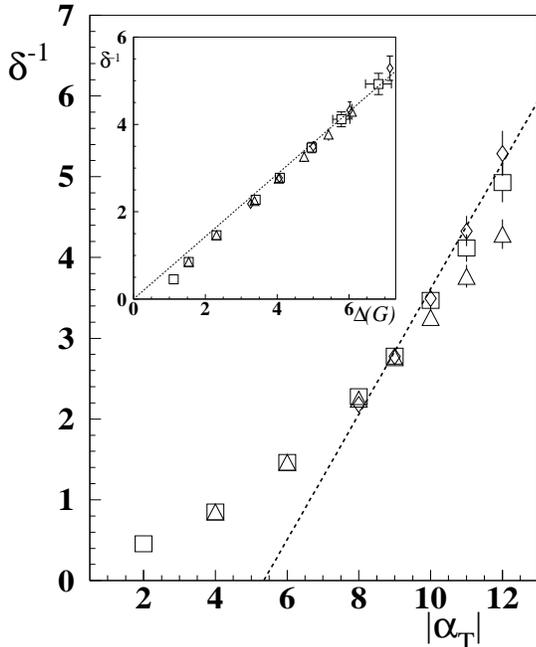}
\\
\caption{
Inverse width of the first peak in $\Delta(k)$ with temperature. The inset 
shows the inverse width against the peak height at different temperatures.
The triangles, squares and diamonds represent $N=132$, $192$ and $402$
respectively. The straight line is a fit to the $N=402$ 
points for $|\alpha_T|\ge8$.
\label{fig:2}}
\end{center}
\end{figure}

We have concentrated our measurements on the form of the peak in $\Delta(k)$ 
at the first reciprocal lattice vector. In our units for a triangular lattice 
with a unit cell of area $\phi_0/B$ this is given by 
$|\hbox{\bf G}|\simeq 2.694$.
In the entire temperature range we look at ($\alpha_T \ge -12$) we find that a
Lorentzian always gives an excellent fit to this peak (see Fig.~\ref{fig:1}). 
This is 
consistent with an exponential decrease of the density correlations in real
space. The 
associated correlation length $\xi_D$ is easily extracted from the width 
$\delta$ of 
the peak at half its maximum, as $\xi_D\propto 1/\delta$. 
In Fig.~\ref{fig:2} we plot $1/\delta$ against 
$\alpha_T$ for different system sizes, which shows how the correlation 
length grows as the temperature is lowered, and reveals when finite-size 
effects become important. 
 Down to $\alpha_T=-10$ the results for the larger systems appear to 
be independent of system size. 
For $\alpha_T<-8$, finite size effects for $N=132$ make its values of 
$\delta^{-1}$ fall beneath the trend of the larger systems.
A linear fit at low temperatures is at least consistent
with the results 
from vortex-vortex correlations\cite{oneill} that $\xi_D \sim |\alpha_T|$,
as $|\alpha_T|\rightarrow\infty$.

For small $|\alpha_T|$
corrections to this scaling form are apparent, so that the range of this fit 
is small. Nevertheless, it is clear that there is no jump in the
value of $\xi_D$ in the vicinity of the observed transition of 
QP simulations, or
the predicted KTHNY transition.
For the integral of $\Delta(k)$ under a given peak to remain 
constant we would expect the general form $\Delta(k)\sim\xi_D f(\xi_D (k-G))$ 
in the vicinity of the peak.\cite{yeo} This implies that the peak height is 
proportional to the inverse width. Using the parameters from our Lorentzian 
fits, this relation is demonstrated in the inset of Fig.~\ref{fig:1}.

These results should be compared to those of Yeo and Moore where $\Delta(k)$ 
was calculated within a high temperature approximation that uses a 
non-perturbative method to go to lower temperatures.\cite{yeo} In this 
work similar 
zero temperature scaling is seen down to $\alpha_T=-10$, although the peak 
heights do not grow as fast. 

In contrast, Hu and MacDonald,\cite{humac} 
who calculate $\Delta(k)$ with an MC 
simulation on a plane with periodic boundary conditions, see an order of 
magnitude jump in the peaks at a transition temperature of $\alpha_T\sim -9$.
 Below this first order transition, $\Delta(k)$ shows ``Bragg peaks'' at the 
reciprocal lattice vectors.
However, in the temperature range above this 
freezing, the correlation function appears to be quantitatively the same as 
on the sphere at the same temperature. This is as it should be as the 
liquid phase on the sphere has the same properties as the liquid on the flat 
plane as long as the system size is sufficiently large.

\subsection{Real Space Phase Correlations}\label{sec:phase}

It is long range coherence in the phase of the  order parameter (ODLRO) 
that gives rise to superconductivity. Within the LLL approximation the phase 
can only be changed by the movement of the vortices, so one might expect a 
simple relation between vortex correlations and phase coherence. Sasik, 
Stroud and Tesanovic\cite{sst} have defined a phase correlation function as 
$\sigma (\hbox{\bf r}-\hbox{\bf r}')=\langle \psi^*(\hbox{\bf r})
\psi(\hbox{\bf r}')\rangle /\overline{\rho}$. 
From measurements of this in their MC simulation 
with QP boundary conditions
 they found that in the thermodynamic limit $\sigma(r)$ fell to zero 
 over the 
order of a lattice constant even when in the solid phase characterized by 
long range density correlations.

As our model is always in the liquid state it is clear that $\sigma(r)$ 
will always be trivially short ranged simply due to the translations of 
the vortex system. However we are interested in the phase
correlations between
 points moving in the same patch of correlated vortices. It is the length 
scale associated with this type of correlation that is physically relevant. 
That is, when this length scale approaches the limiting scale of a given 
sample (eg. the distance between pinning centers) it will become 
superconducting. To investigate this length scale we need to 
measure the correlations in the phase from a fixed vortex. We also need 
to fix the rotations about this vortex (these would also destroy phase 
coherence). In practise we can do this by (a) fixing one vortex at a pole
($\theta=0$) 
by setting $u_0=0$, 
and (b) fixing the nearest six vortices by keeping $u_i=u_i^{(0)}$ for 
$i=1,2,\ldots 6$. 
$\{u_i^{(0)}\}$ are the coefficients for the ground state with a 
fixed orientation.
These conditions effectively fix the order parameter in a local region 
around  the central fixed vortex. 
We can then measure the correlation between the phase at a point 
$\hbox{\bf r}'=(\theta',\phi')$ 
and at $\hbox{\bf r}=(0,0)$:
\begin{equation}
\sigma_{fix}(\hbox{\bf r}')=
\langle \psi^*(\hbox{\bf 0}) \psi(\hbox{\bf r}')/
| \psi^*(\hbox{\bf 0})|| \psi(\hbox{\bf r}')| \rangle_{fix} .
\end{equation}
Note the difference between this and the correlation function of SST. 
As well as the fixing of the vortices, our definition 
contains no part dependent on the amplitude of the order parameter. 
In terms of the MC variables, the phase at $\theta=0$ is given by
$\psi(0,\phi)/|\psi(0,\phi)|=u_1e^{i\phi}/|u_1|$ where $\phi$ is the 
azimuthal angle of approach as $\theta\rightarrow 0$. Fixing this angle $\phi$
to be 
zero, the phase correlation function may be written as
\begin{eqnarray}
\sigma_{fix}(\theta',\phi')=&&\nonumber\\
&&\hspace{-2cm}
\left\langle \frac
{u_1^* \sum_{m=1}^N u_m h_m \sin^{m}
(\theta'/2)\cos^{N-m}(\theta'/2)e^{im\phi'}}
{ \left| u_1^* \sum_{m=1}^N u_m h_m \sin^{m}(\theta'/2)\cos^{N-m}(\theta'/2)
e^{im\phi'}  \right|}
 \right\rangle_{fix} .
\end{eqnarray}

The reader may notice a problem in measuring the phase of a 
complex function on the surface of a sphere. If a path is followed that 
encircles one vortex, the phase of $\psi$ will change by $2\pi$. However, 
an ambiguity exists in the definition of the inside or the outside of a 
closed loop in a closed surface and this could lead to an ill defined phase. 
In fact, it is our choice of gauge that 
places on  one axis a string (or infinitely long and thin solenoid) 
terminating  at the monopole that removes this ambiguity. 
This choice effectively makes the South 
pole ($\theta=\pi$) the ``outside'' of the system, and the North pole 
($\theta=0$) the center. This results in a singularity in the phase of the 
order parameter at $\theta=\pi$.  
Encircling the South pole 
\end{multicols}
\widetext
\begin{figure}[tbp]           
\epsfxsize= 17.5cm
\begin{center}
%\leavevmode\epsfbox{figmd7.eps}
\caption{
The top diagrams show the phase correlation function 
$|\sigma_{fix}(\theta,\phi)|$ plotted in the
complex $z$-plane where $z=\tan{(\theta/2)}e^{i\phi}$
for $\alpha_T=-6, -9$ and $-11$, with $N=72$. The bottom row shows the mean 
density $\langle {|\psi(\hbox{\bf r})|}^2\rangle_{fix}/\overline{\rho}$ 
with the same 
fixed coefficients as for the phase correlations. 
Notice that one of the fixed
 vortices is a five-fold center.
\label{fig:4}}
\end{center}
\end{figure}
\begin{multicols}{2}
without apparently 
encircling any vortex results in a phase 
change of $2N\pi$ (this is the origin of the Dirac quantization of the 
charge on a monopole). This does not imply a singularity in the supercurrent, 
as the gauge part of the current will exactly cancel with this wave function 
contribution.

\begin{figure}[htbp]           
\epsfxsize= 8cm
\begin{center}
\leavevmode\epsfbox{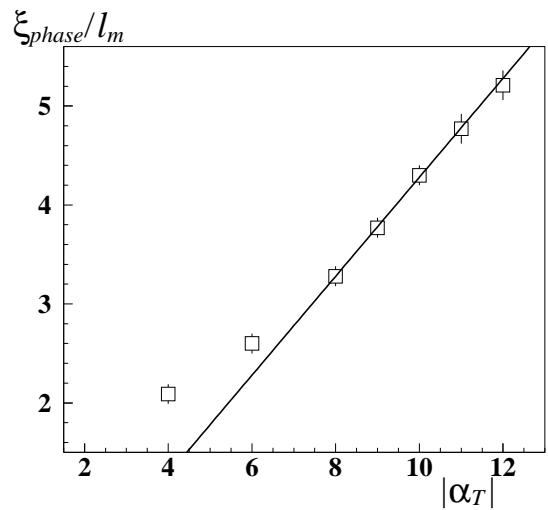}
\vspace{5mm}
\caption{
The length scale associated with phase coherence extracted from 
$\sigma_{fix}(\theta,\phi)$ with $N=192$ against temperature.
 \label{fig:6}}
\end{center}
\end{figure}

Our measurements of $\sigma_{fix}(\theta,\phi)$ are shown in Fig.~\ref{fig:4}
 using 
the projection $z=\tan{(\theta/2)}e^{i\phi}$. 
For comparison, we also show the mean density 
$\langle{|\psi(\hbox{\bf r})|}^2\rangle_{fix}$ with the same coefficients 
fixed.
A growing range of phase 
coherence as the temperature is lowered is demonstrated, and coincides 
with the growth in density correlations.  
The length scale $\xi_{phase}$ is found by plotting $\sigma_{fix}$ against 
$R\theta$ and fitting a gaussian over the region outside the fixed vortices, 
i.e.\ for $R\theta>a$ with $a=2.5$.  We find that  $\xi_{phase}$ 
 increases proportionally to $|\alpha_T|$ in the temperature 
range $\alpha_T<-7$ (see Fig.~\ref{fig:6}), with again corrections to 
scaling visible for $\alpha_T\gtrsim -7$. 
The similarity between Figures~\ref{fig:2}~and~\ref{fig:6} indicates that 
the length scales 
for phase and density coherence grow proportionally to each other as the 
temperature is reduced.
The length scales are unchanged within 
the measured errors for the different system sizes studied. 
A relation between the growth of 
correlations in the density ${|\psi|}^2$ and in the phase of $\psi$ is 
clearly observed.

The scaling of phase coherence $\xi_{phase}\sim|\alpha_T|l_m$ was predicted by 
O'Neill and Moore 
from two distinct arguments. One of these involved
considering the thermal excitation of small amplitude shear modes from the 
ground state 
lattice  to define a phase coherence length within the crystalline phase. 
The second  
argument considered the free energy cost of phase fluctuations over a 
finite region of the vortex system. It is this second argument that is 
directly relevant to the liquid phase with a finite length scale of 
crystalline order,
and we repeat and amplify it here. 
The free energy cost of 
phase fluctuations over a region with crystalline correlations
was shown to be 
$F_{eff}=(d/2)\int d^2rc_{66}l_m^4(\nabla_{\perp}^2\Phi)^2$,
where $\Phi$ is the phase of the order parameter. From this result, the free 
energy cost of a phase change of $\pi$ over a region of linear extent $\xi$
will be $F(\xi)\sim c_{66}dl_m^4\xi^{-2}$. 
When such excitations proliferate, 
phase coherence will be destroyed. Hence, there will be phase coherence only 
over the length scale $\xi$ when the free energy $F(\xi)$ is of order $k_BT$.
The  $\alpha_T$ dependence of the mean-field elastic shear modulus $c_{66}$ 
is known 
in the large field limit, $c_{66}\simeq \alpha_T^2k_BT/dl_m^2$, with the 
simple result that $\xi\sim|\alpha_T|l_m$.

\narrowtext

\begin{figure}[htbp]           
\epsfxsize= 6cm
\begin{center}
\leavevmode\epsfbox{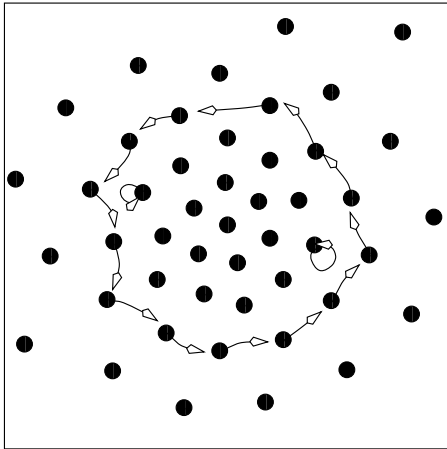}
\caption{
The path of vortices under the change 
$u_m\rightarrow u_m'=u_me^{i\gamma}$ for $m=0,1,2,\ldots\nu$ as $\gamma$ goes from $0$ to $2\pi$ with $\nu=20$. In this case the original state is the 
ground state for $N=72$. The vortices are plotted in the 
complex $z$-plane where $z=\tan{(\theta/2)}e^{i\phi}$.
\label{fig:braid}}
\end{center}
\end{figure}

We believe that there is another
 kind of phase fluctuation within the liquid  that can be 
identified with a topological defect-- a ``braid''. Consider
 a ring of vortices of a given radius that move 
around the circumference of the ring. (In three dimensions this vortex 
movement may take place along the field direction to form a topologically 
constrained defect-- hence the name, braid. We have previously calculated the 
energies of such 
defects.\cite{dodgson2})  Such a movement may be created by a simple 
combination of changes to the basis coefficients. If we make the transformation
$u_m\rightarrow u_m'=u_me^{i\gamma}$ for $m=0,1,2,\ldots\nu$ and make no 
change to the 
remaining coefficients $m=\nu+1,\ldots N$ then such a braid movement will 
occur in a 
ring of vortices that enclose approximately $\nu$ vortices 
(see Fig.~\ref{fig:braid}). 
If $\gamma =2\pi$ the vortices 
in the ring will each hop to a neighbors position, with a phase change in 
the order parameter of $2\pi$ inside the ring (i.e.\ no essential change). 
Excitations will involve values of $\gamma$ less than $2\pi$ which increase 
the phase at the center of the ring by $\gamma$ while leaving the phase 
unchanged outside of the ring. The energies of these braid excitations will 
define a length scale over which they proliferate, and thus destroy phase 
coherence. The energy costs of braids 
in a crystal phase would 
increase with the size of the ring, and so 
phase coherence would not be destroyed by this mechanism in the crystalline
state. Nevertheless, we believe that the excitations of
such defects are 
present on the length scale $\xi$ in the liquid phase, and are
a plausible mechanism for the loss of phase
coherence seen in Fig.~\ref{fig:4}. Notice that in Fig.~\ref{fig:4} it is when
the density correlations become ring-like (as would happen due to braid 
excitations) that the phase coherence is lost.

\section{Effect of an inhomogeneous magnetic field}\label{sec:inhom}

If an experiment using the spherical geometry was ever to be realized, 
there would always be a problem in placing the end of a solenoid 
(the ``monopole'') exactly at the center of the sphere. This will 
cause the external field to be inhomogeneous over the surface of the 
sphere. Such inhomogeneities are generic in experiments with any 
geometry, so it is worthwhile investigating these effects.

We consider displacing the monopole from the center of the sphere by a 
distance $d$ along the $z$-axis. It is straightforward to show that 
for small $\epsilon=d/R$, this induces a change in the vector potential of 
$\delta\hbox{\bf A}=\epsilon B_0 R\sin{\theta}\hat{\hbox{\boldmath$\phi$}}$.
An inhomogeneity is introduced into the magnetic field at the surface which
has a radial component of $B_r=B_0(1+2\epsilon\cos{\theta})$.
Substituting this new field into the free energy gives an extra term to 
first order in $\epsilon$ of
\begin{equation}
\delta F=\frac{\tilde{\epsilon}}{N}\sum_{m=0}^N|v_m|^2(2m-N),\label{eq:extra}
\end{equation}
where $\tilde{\epsilon}=(d\hbar^2Q^2/2m)\epsilon$. 

Before we look at our MC results with this extra term, we consider
its effect within the high-temperature perturbation theory. With the addition 
of (\ref{eq:extra}) to the Hamiltonian, the HF equation becomes
\begin{equation}
\tilde{g}_m=g_m(1-2\tilde{g}_m\sum_p\tilde{g}_p\beta_{pmpm}),
\end{equation}
with $g_m=\alpha_T+\tilde{\epsilon}(2m-N)/N$. The sum in this equation is no
longer independent of $m$. However, in the large $N$ limit,  $\beta_{pmpm}$
has the form,
\begin{equation}
\lim_{N\rightarrow\infty}(\beta_{pmpm})=
\frac{1}{2\Delta\sqrt{\pi}}\exp[-(p-q)^2/\Delta^2],
\end{equation}
where $\Delta=[(2N-p-q)(p+q)/4N]^{1/2}$. The width of the interaction is
of order $N^{1/2}$ over most of the sphere. 
If the variation in the renormalized propagator $\tilde{g}_p$
is smooth over the range of  $N^{1/2}$ (which corresponds to changes in real
space being smooth over the distance $l_m$) then the sum over $p$
is dominated by contributions when $p\simeq m$ to give
$\sum_p\tilde{g}_p\beta_{pmpm}\simeq\tilde{g}_m/2 $. 
Therefore we have $N+1$ independent HF 
equations $\tilde{g}_m=g_m(1-\tilde{g}_m^2)$. Each has the same solution,
 but with an $m$-dependent effective temperature:
$\tilde{g}_m=(\alpha_m/2)[-1\pm(1+4/\alpha_m^2)^{1/2}]$ with
$\alpha_m=\alpha_T+\tilde{\epsilon}(2m-N)/N$.
This will be true for any change in the Hamiltonian of the form
$\delta F=\sum_{m=0}^N\epsilon(m)|v_m|^2$. where the effective temperature
becomes $\alpha_m=\alpha_T+\epsilon(m)$.
Such a change corresponds to changes in real space in one direction only
over length scales longer than $l_m$.

This procedure can be generalized to an arbitrary diagram in the 
perturbative scheme.
Any diagram is dominated by contributions where the indices in and 
out of a vertex are within $\Delta$ of each other. Therefore a given 
diagram is just the sum of
the respective diagrams without the inhomogeneity but at the effective 
temperature $\alpha_m$.
For example, the free energy may be expanded in a power series of the 
propagator, $F_{pure}(\alpha_T)=N\sum_{n=0}^\infty a_n\tilde{g}^{2n}$. 
With the 
inhomogeneity this generalizes to
\begin{equation}
F_{inhom}(\alpha_T,\tilde{\epsilon})=
\sum_{n=0}^\infty\sum_{p=0}^N a_n\tilde{g}_p^{2n}=
\frac{1}{N+1}\sum_{p=0}^NF_{pure}(\alpha_p).
\end{equation}
This result may break down when variation in
$\epsilon(m)$ is on a scale less than $N^{1/2}$. 
 In the thermodynamic limit
our conclusion will hold as long as the perturbation expansion 
remains useful.

\begin{figure}[htbp]           
\epsfxsize= 8cm
\begin{center}
\leavevmode\epsfbox{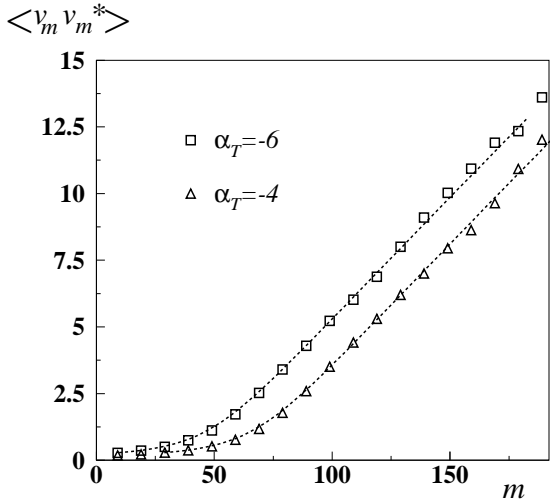}
\\
\caption{The $m$ dependent propagator for two different strengths of
inhomogeneity with $N=192$ and
$\tilde{\epsilon}=-10$. 
The dashed line represents the value of the
propagator in the pure case at the effective temperature 
$\alpha_T=\alpha_m$. Note that the effective temperature covers the range
 $-16 < \alpha_m < +4$.
\label{fig:homvs}}
\end{center}
\end{figure}
Interesting effects may occur when there is a phase transition in the 
pure system. However, as our simulations fail to see such a transition,
the above analysis is sufficient to explain our numerical results. 
In Fig.~\ref{fig:homvs} 
we plot the $m$-dependent propagator $|v_m|^2$
from MC simulations with $\tilde{\epsilon}=-10$. and compare to the
zero inhomogeneity propagator with the $m$-dependent effective temperature.
The results confirm that the system behaves with the local properties of the
effective temperature.

\section{Discussion of simulations with different geometries}
\label{sec:discussion}

In this section we discuss some of the problems of 
different boundary conditions when interpreting the results of numerical
simulations of superconducting thin films within the LLL approximation.
We first mention some of the shortcomings 
of the spherical 
geometry used in our simulations. We then go on to discuss the other 
main geometry used-- QP boundary 
conditions. 
In this case there exist problems not present in the 
spherical geometry due to the constraints on the phase of the
order parameter. 
We also consider a third geometry with a circular geometry, where initial
work has not found a phase transition.
Finally, we look at the implications of
recent experiments.

\subsection{The spherical geometry}

An important difficulty on the sphere is that the crystalline
 state free energy is
raised due to the curvature and the necessary presence 
of twelve disclinations. 
A liquid with correlation 
length less than the system size will be unaffected by the curvature.
The energy difference should disappear 
in the thermodynamic limit (see Section~\ref{sec:ground}). However, 
for the system sizes 
studied, this energy remains on the same order of magnitude of the jump in 
energy seen at the transition observed with QP boundary conditions. 

The presence of the topologically required disclinations on the sphere 
is clearly an important feature not present in real thin films. 
However, their effects will only become important as the correlation 
length of the system approaches the distance between disclinations. 
This is a finite size effect as experienced by any geometry, although 
the effective system size will be smaller for a given $N$ than on a plane. 
For the largest system sizes studied, finite-size effects 
probably  become important in the region of interest at $\alpha_T \sim -10$.

The mobility of these disclinations, if large, would be a mechanism
for destroying crystalline order in a finite-sized system. At
low temperatures the disclinations should not be mobile, 
both because of their mutual repulsion (as can be seen from the 
elasticity calculation) and the topological barrier to their motion 
through the lattice. At higher temperatures where there is a finite 
density of dislocations (eg. in a hexatic phase) the dislocations 
provide a topological mechanism for the motion of the disclinations.
In finite size systems, the strain field associated with the disclination 
may even create dislocations by pulling apart the dislocation pairs that 
are always present in a 2D crystal. From inspection of snapshots of the 
systems at finite temperatures we have observed that dislocations are
attracted to the disclinations, and that the positions of the disclinations
are not perfectly icosahedral as they are in the ground states.

Our measurements so far have been insensitive to hexatic bond-angle order.
This order could never be globally defined on the sphere,
and will only be a meaningful concept within the regions free of 
any disclinations, which  are small in the system sizes we study.

It must be conceded, therefore, that our numerical evidence 
from the sphere for the non-existence 
of a phase-transition is not yet conclusive. Larger system size studies 
are clearly needed, but even increasing sizes by an order of magnitude
may not resolve the problem. Our work on the ground states suggests that very
large systems with $N\sim 10^4$ may be needed to reach certain properties
of the thermodynamic limit such as the screening of the disclinations by
dislocations.

\subsection{Quasi-periodic boundary conditions}

In the system investigated by 
Refs.~\onlinecite{tesanovic,kato,humac,sasik,sst} 
a constant magnetic field {\bf H} 
perpendicular to a thin-film superconductor is represented by the Landau 
gauge, $\hbox{\bf A}=-By\hat{\hbox{\bf x}}$. In this gauge the LLL states are
$\psi_p(x,y)=\exp{[ipk_0x-(y-pk_0l_m^2)/2l_m^2]}$, where $k_0$ is a 
``momentum'' in the $x$ direction that must be fixed by the boundary 
conditions.
For a general state in 
the LLL subspace, $\psi(x,y)=Q\sum_p c_p \psi_p(x,y)$, the system is 
periodic in the x direction with period $L_x=2\pi/k_0$. Quasi-periodicity 
in the y-direction may also be imposed by setting $c_{m+N}=c_m$. This 
condition leads to $\psi(x,y+L_y)=\exp(iNk_0x)\psi(x,y)$ with $L_y=Nk_0l_m^2$.
So each principle region has an area $L_xL_y=2\pi Nl_m^2$ and contains $N$ 
zeros of $\psi$ that correspond to vortices. To have a geometry 
commensurate with a triangular vortex lattice, 
$N_x=L_x/l_0$ and $N_y=L_y/(\sqrt{3}l_0/2)$ are fixed as integers.
(Note the difference between the magnetic length $l_m$ and the lattice spacing
$l_0$ both of which are defined in Section~\ref{sec:method}.)

The difference between this system and more conventional periodic boundary 
conditions comes from the constraints on the phase of the order parameter 
as the principal region is crossed. That such a constraint exists is 
apparent from the fact that there are only $2N$ degrees of freedom available 
to describe the $2N$ coordinates of the vortex positions as well as an 
overall phase and amplitude. It has been shown that the phase constraints 
lead to the center of mass of the vortex positions to be fixed,\cite{rezayi} 
which 
accounts for the missing two parameters.
A consequence of the fixed center of mass is that there are only $N$ 
distinct ground states, each a triangular lattice separated by a minimum 
discrete distance.

In MC simulations with QP boundary conditions, evidence is seen for
 a first order transition. The observed melting temperature drops as 
system size is increased but extrapolation gives an estimate for the 
thermodynamic transition\cite{humac} 
at $\alpha_T=-9.3\pm 0.1$.
In the low temperature state dominated by the ground state there is a finite 
energy barrier between different ground states. We now estimate what that 
energy barrier is.

\begin{figure}[htbp]           
\epsfxsize= 8cm
\begin{center}
\leavevmode\epsfbox{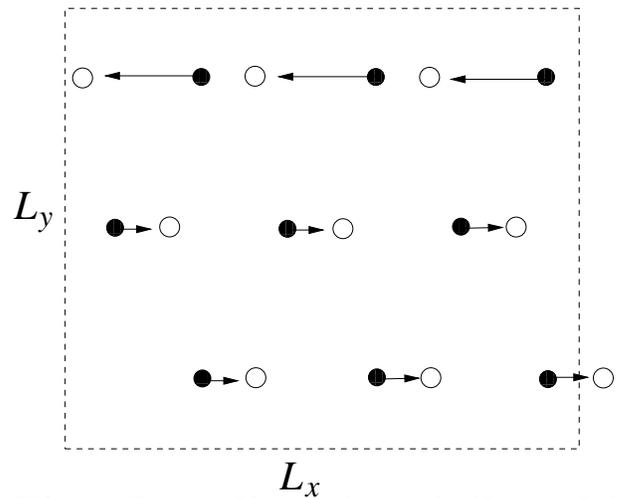}
\\
\caption{
Two possible ground states for $N=9$ with QP boundary conditions. A path 
between the two ground states is shown that leaves the position of
the center of mass unchanged. 
Note that only the principal region is shown.
\label{fig:qp9}}
\end{center}
\end{figure}

Figure~\ref{fig:qp9} 
illustrates the sort of movement of vortices between ground 
states that obeys the center of mass constraint. The example is for $N=9$, 
and a short inspection will reveal that there are only 9 distinct ground 
states that may be related by such moves. In general the different ground 
states have the form $c_{p-q_1}=\exp{[i(\pi/2)(p/N_x)^2-2i\pi(p+N_xq_1)q_2/N]}$
for $p$ equal to an integer multiple of $N_x$, and zero otherwise. Each 
ground state is labeled by $\psi_{q_1,q_2}$, where $q_1$ 
must be an integer from $1$ to $N_x$, and 
the condition $c_p=c_{p+N}$ quantizes the allowed values of
$q_2$ to integers from $1$ to $N_y$. These ground states are exactly 
equivalent to the Eilenberger\cite{eilenberger}
 states $\phi(\hbox{\bf r}|\hbox{\bf r}_0)$, 
with $ \hbox{\bf r}_0=(x_0,y_0)$ where $x_0=q_2l_0/N_y$ and
$y_0=q_1(\sqrt{3}/2)l_0/N_x$. These are related by 
$\phi(\hbox{\bf r}|\hbox{\bf r}_0)=\exp{[2\pi i(2y_0/\sqrt{3})x]}
\phi(\hbox{\bf r}+\hbox{\bf r}_0|
\hbox{\bf 0})$.

Consider the change $c_p\rightarrow c_p'=c_pe^{ip\gamma}$ for any 
configuration of $\{ c_p\}$.
For an infinite system this just corresponds to a uniform translation of
the vortices in  the x direction by $\gamma/k_0$. However, with QP boundary 
conditions this change will involve movement of vortices at the edges of the 
principal region in the opposite direction, thus conserving the center of mass.
(An analogous change in the spherical geometry gives the braid in
Section~\ref{sec:phase}.)
It is exactly this sort of movement that is a path between different ground 
states. For example, if we start from $\psi_{0,0}$ and apply this 
transformation with $\gamma$ changing from $0$ to $2 \pi q_2/N$ we arrive at 
the state $\psi_{0,q_2}$.

What is the energy barrier for such a path? The free energy of the QP 
system is given by
\begin{eqnarray}  
{\cal H}\left( \{c_m\}\right) =k_B T&&
\alpha_T^2\Biggl[\hbox{sgn}{(\alpha_T)} \sum_{m=0}^{N-1} |c_m|^2
+\nonumber\\
&&\frac{1}{2N_x}\sum_{p=0}^{N-1}\sum_{q,r,s=-\infty}^{\infty}
 \beta_{pqrs}c_pc_qc_r^*c_s^*\Biggr]  ,
\end{eqnarray}  
with $\beta_{pqrs}=3^{1/4}2^{-1/2}
\delta_{p+q,r+s}e^{\{
-[(p-q)^2
+(r-s)^2]\sqrt{3}\pi/4N_x^2\}}$.
Changing the phase of the coefficients does not affect the quadratic energy. 
Also, the contributions to the quartic term $\beta_{pqrs}c_pc_qc_r^*c_s^*$ 
will be zero when $p$, $q$, $r$, and $s$ correspond to the 
same region (ie. they are all between $0$ and $N-1$). 
As the width of the gaussian in the quartic interaction is
of order 
$N_x$, contributions will be negligible unless these variables  are 
within $\sim N_x$ of each other, and the edge of the principal region. 
Therefore the energy barrier of this path 
will be of order $N^{1/2}$. This $N$ 
dependence has been observed 
numerically.\cite{sasik2}

\begin{figure}[htbp]           
\epsfxsize= 8cm
\begin{center}
\leavevmode\epsfbox{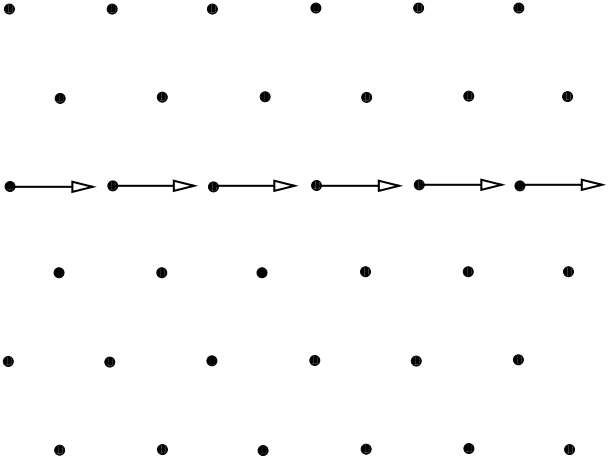}
\\
\caption{
The relative motion of vortices when moving between two neighboring
ground states in the QP system.
\label{fig:inf}}
\end{center}
\end{figure}

A more quantitative estimate may be obtained by considering the movement 
of the vortices at the edge of the principal region relative to the 
vortices in the bulk of this region (see Fig~\ref{fig:inf}). 
At any point on the path this represents a
defect from the perfect ground state, the energy cost of which
may be calculated using the methods of 
Ref.~\onlinecite{dodgson2}. We find an energy barrier of
\begin{equation}
E_b=0.33 (L_x/l_0)\alpha_T^2,
\end{equation}
which is proportional to $N_x$ as predicted.
This is an upper bound, as only the movement of a line of vortices 
has been allowed, without relaxation of the surrounding lattice. However, 
as the interaction energy is short ranged,\cite{dodgson2} we do not 
expect the $N$ dependence to change. It is also a zero temperature result, 
but it should still hold for a low temperature state with long correlations.
We emphasize that this energy barrier grows with system size, despite the fact 
that the positions of the discrete ground states get closer to each other.

Additional evidence for the size of these barriers comes from the 
simulations themselves. In Ref.~\onlinecite{sst} a plot is shown of 
$\langle |\psi(\hbox{\bf r})|^2\rangle$ at a temperature corresponding 
to the extrapolated thermodynamic transition, where the modulations in 
this quantity show that the system is dominated by just one of the $N$ 
ground states during the length of the measurement, $10^6$ MC steps.

The measurements by SST of this low temperature 
state are clearly not equilibrated as not all of the ground states are being
sampled; there should always be a translationally invariant density for
any {\em finite} system.
It is also a possibility that energy barriers between ground states 
affect the evidence 
for a first order phase transition. For instance, the free energy of the 
liquid will be overestimated if the simulation is not run for
long enough (the standard time for these simulations has been 
$\sim 10^6$ MC steps) which may alter the melting temperature.  
This casts doubt on the extrapolation of the melting temperature to the 
thermodynamic limit.
Even if a first order (or continuous) transition remains after full 
equilibration, the discrete ground state problem makes the QP system 
unsuitable for transport measurements.

\subsection{The circular geometry on a flat plane}

An alternative set of boundary conditions may be 
constructed in the symmetric 
gauge,\cite{oneill} $\hbox{\bf A}=
(x\hat{\hbox{\bf y}}-y\hat{\hbox{\bf x}})B/2$. The LLL states
may be formed by the complete basis with
circular symmetry $\psi_m=\zeta^m\exp[-|\zeta|^2/(4l_m^2)]$,
where $\zeta=x+iy$. To perform numerical simulations, one has to
truncate the series of these states to a finite number of terms, say,
$m=0$ to  $N$.
This truncated system has $N$ vortices, as $\psi=\sum_{m=0}^N c_m\psi_m=
\exp{[-|\zeta|^2/(4l_m^2)]}\prod_{i=1}^N 
(\zeta-\zeta_i)$, and the order parameter
drops rapidly to zero outside of a radius $(2N)^{1/2}l_m$ from the origin.

Little work has been done on this geometry as finite size effects 
are strong. Numerical minimizations on small systems have shown 
that the 
ground state energy per vortex is {\em lower} 
than for the infinite plane,\cite{oneill2} and vortices near the edge of the
system are attracted to the ``boundary''\cite{sasik2} 
(near where the order parameter is zero).
The finite size effects are greater than with spherical or QP boundary
conditions where there are no edges of the finite systems and each vortex
resides in an equivalent local environment. The
effect of the boundary is mainly in a fraction of $\sim N^{-1/2}$ of
the bulk of the system, and should be unimportant in the
 thermodynamic limit.

\begin{figure}[htbp]           
\epsfxsize= 8cm
\begin{center}
\leavevmode\epsfbox{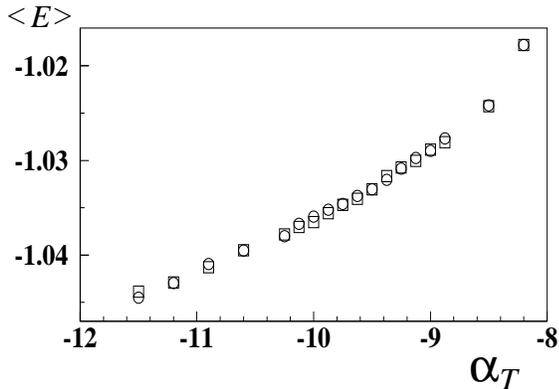}
\\
\caption{
The mean internal energy from MC measurements on the 
circular geometry with $N=1000$. The circles represent measurements 
while the temperature is being raised; the squares represent cooling. Each 
measurement is averaged over 30 000 MC steps. The energy is normalized
by the ground state energy per vortex on the infinite plane.
\label{fig:circ}}
\end{center}
\end{figure}

A similar geometry has been used in simulations of the 2D one component 
plasma (OCP), where the particles are confined to a circular disk.\cite{choq}
A first order transition was observed and this contrasts with the 
absence of a transition in simulations of the OCP on a spherical 
geometry.\cite{caillol}
This has motivated us to perform simulations on the circular geometry
in search of a phase transition. We have performed the real space phase
and density correlation measurements as in Section~\ref{sec:phase} for systems
up to $N=400$ and find results consistent with those on the sphere. We
have also measured the internal energy while heating from the ground state
and cooling from high temperatures, for system sizes up to $N=1000$. These
energy measurements (see Fig~\ref{fig:circ})
have revealed no hysteresis on the scale observed in the QP
system for temperatures greater than $\alpha_T=-12$, and hence 
 no evidence for a phase transition.

\subsection{Experiments}

It seems that from a theoretical point of view there are three possible 
scenarios for the behavior of thin film superconductors in the 
presence of a magnetic field:
a) the standard picture of KTHNY melting of a 2D crystal where a dislocation
mediated continuous transition occurs between a low temperature state 
with quasi-crystalline order and a high temperature liquid or hexatic phase;
b) a first order transition where the liquid free energy becomes
 lower than the solid phase before KTHNY melting takes place (as is seen in 
QP simulations); c) the absence of a distinct low-temperature state, 
but where the correlation lengths of the liquid diverge as $T\rightarrow 0$ 
(this is the prediction of high order perturbation theory from the high 
temperature side, and is consistent with the numerical simulations on 
the sphere).
We now discuss some experimental results and their implications for 
these possibilities.

A melting transition of the flux lattice in In/In0$_x$ thin films was 
observed by Gammel {\it et al.}\cite{gammel} 
The zero field limit of the transition was 
consistent with the Kosterlitz-Thouless transition in the absence of 
external fields, 
though the magnetic field dependence of the melting line did not 
satisfy the predictions of the KTHNY theory. The measured transitions were 
mainly at low fields, so the relevance of our simulations here are doubtful.

In the experiments of Berghuis {\it et al.}\cite{berghuis} the current voltage 
response was measured of amorphous Nb$_{1-x}$Ge$_x$ films at different 
fields and temperatures. A drop to zero 
in the a.c.\ ohmic resistivity is seen, for 
example at fields $H\sim 0.8
H_{c2}$ at $T=0.68T_c$, though the d.c.\ response remains finite 
(ie. the critical current is zero.). This drop is attributed to a KTHNY 
transition consistent with the temperature and film thickness dependence 
of the melting field.
A similar conclusion is made by Yazdani {\it et al.}\cite{yaz} from a.c. 
measurements on films of amorphous Mo$_{77}$Ge$_{23}$. 

It should be 
emphasized that these a.c.\ measurements at a given frequency $\omega$
probe structural changes of the vortex system  over a restricted
length scale.\cite{yaz} While this is useful 
to measure properties independent of the presence of  a large
pinning length, one would expect even in the absence of a phase transition 
to see a large drop in the resistivity at fixed frequency
 when the relaxation times grow larger than $1/\omega$.
In numerical simulations on the sphere,\cite{hanlee} 
the plastic time scale was observed 
to increase as $\tau_{pl}\sim\exp{(A\alpha_T^3)}$ and we would expect a sharp 
drop
when $\tau_{pl}\sim 1/\omega$.

In contrast to the above experiments, measurements by Nikulov 
{\it et al.}\cite{nikulov} on films of
amorphous NbO$_x$ with weak pinning have found no signatures of a 
resistive transition down to fields $\sim 10^{-3}H_{c2}$. This implies 
that the transitions observed previously are not universal properties of 
superconducting thin films. While the lack of a resistive drop may be 
attributed to the total absence of pinning, one would expect in the 
presence of a first order transition, where there is a discontinuous 
change in $\langle |\psi|^2\rangle$, that this would be observed as a 
step in the flux flow resistivity.

Thus the 
interpretation of these experiments remains controversial. While the KTHNY 
theory provides a useful framework to discuss flux lattice melting in two 
dimensions, its effects have only been observed over short length 
scales.\cite{yaz} 
As far as we know there has been no experimental indication of a first 
order transition in superconducting thin films. In principle, the results 
of these experiments can still 
be explained simply by the growing length scales 
and relaxation times of a vortex liquid in the absence of a phase transition.

\section{Conclusions}\label{sec:conclusions}

We have shown that the model with a spherical geometry remains in the vortex 
liquid phase, at least down to temperatures $\alpha_T\gtrsim -12$
for system sizes up to $N\le 402$. 
This could be due 
to the raised energy of the ground states because of the necessary presence 
of twelve disclinations in a triangular network over a sphere. However, we 
have made similar investigations on an alternative finite size model: 
considering the LLL basis states on an infinite plane with a circular 
symmetric gauge and truncating the series to include only $N$ 
vortices.\cite{oneill} This system has a lower ground state energy per vortex 
than the infinite system and no disclinations, but we see the same length 
scales as in Section~\ref{sec:phase} for temperatures down to $\alpha_T=-12$. 
The main evidence for a first order transition in the QP model is the 
double peak structure observed in the probability distribution of the 
internal energy at the transition. Our investigations on the internal 
energy for both the sphere and the plane with circular coordinates have
revealed no  such signatures.

The fact that the  spherical system remains in the liquid phase down to low 
temperatures allows us to measure the variation of the correlations in the 
liquid state with temperature over a large regime. 
Measurements in the liquid phase have the 
advantages that the system reaches 
equilibrium over a reasonable time scale at low temperatures. 
Our measurements of the density correlations  are 
consistent with those in the liquid phase using QP boundary 
conditions.\cite{humac} The extracted 
length scales demonstrate an asymptotic linear dependence on $|\alpha_T|$. 
The scaling analysis of O'Neill and Moore predicted that the {\em phase 
coherence} length should grow as $\xi_{phase}\sim |\alpha_T|$. 
Our results for the 
phase coherence are consistent with this scaling form, and also demonstrate 
that in the liquid the phase and density correlations grow with identical  
temperature dependence.

Our results for the scaling of the phase correlations 
contradict the conclusions of SST, 
who claim that in the thermodynamic limit $N\rightarrow\infty$ the phase 
coherence has a range only of the order of the lattice spacing 
in their ``solid'' 
phase. 
However, this discrepancy is really due to different 
definitions of what is meant by phase coherence.
SST do not predict the growth of the length scale of phase correlations 
as the temperature goes to zero. Although their results are claimed as
evidence for the charge-density wave  of Tesanovic,\cite{tesan} 
a finite length scale 
that grows with decreasing temperature or even a power law form of ODLRO
would still be consistent with this exotic phase.
It seems unlikely to us that 
the low-temperature phase below a first order transition
would have a 
shorter range of phase coherence than  the liquid phase.  
If we measure the phase 
correlations without fixing any vortices, we also find no phase coherence 
beyond a length of one lattice spacing, due to the translational freedom of 
the liquid phase. In the solid phase of the QP model 
the vortex system appears to be 
fixed near one ground state due to the finite energy barrier between the $N$ 
ground states, and this allows phase coherence to be observed 
over a long 
range. As $N$ increases, there is a decrease in the range of phase coherence.
These measurements by SST demonstrate just how strong the finite-size
effects are in the QP simulations below the observed transition.
Without fixing the motion of any vortices, the phase coherence should always 
have a
trivial length scale of the inter-vortex spacing.
By restricting the motion of a small number of 
vortices, we have been able to measure a more physically relevant phase 
correlation length that is consistent with the expected growth of order 
as the temperature is reduced. An interesting question remains as to whether 
a similarly defined phase coherence in the solid phase with QP 
boundary conditions would result in 
true long range order, or another length scale that also exhibits 
zero temperature scaling.

Despite the differences in results from spherical and  QP boundary 
conditions, our results from high temperature perturbation theory and 
on the ground states suggest that the thermodynamic limit will be the 
same on the sphere as the infinite flat plane.
This contradicts the earlier statement of Ref.~\onlinecite{dodgson} where the 
effect of allowing additional dislocation defects 
into the ground state was not considered. 
Even so, the
results on the ground states show how large system sizes must be for 
certain properties to approach the thermodynamic limit. A definite 
drawback on the sphere is the small effective system size compared to 
the number of degrees of freedom.

The spherical geometry is perhaps not only 
relevant for simulational purposes. It is possible that an experiment could be
constructed with an amorphous type-II superconductor sputtered over the 
surface of a sphere, and one end of a long and thin solenoid inserted 
into the center  to approximate the field from a monopole. In contrast
an experimental realization of the QP boundary conditions is not possible
(these boundary conditions are primarily of interest in extrapolating 
to the thermodynamic limit of 
an infinite number of free vortices). 
We feel that the spherical geometry,
with its property of translational invariance, 
 remains a useful system for 
numerical simulations, especially to measure properties of the unpinned
vortex liquid.
An investigation into dynamical properties with a view to examining the 
length scales associated with non-local resistivity\cite{blum} 
is planned, which 
would not be possible using QP boundary conditions
due to 
the absence of full translational invariance
of the vortex positions.

%\newpage
%\begin{center}
%{\bf ACKNOWLEDGEMENTS}
%\end{center}
%\bigskip
\acknowledgements

We would like to acknowledge useful interactions with 
Gianni Blatter, Tom Blum, Vadim Geshkenbein,
Roman Sasik
and Joonhyun Yeo. MJWD would
also like to thank Han Lee for the use of his source code and the EPSRC 
for financial support.

\appendix 

\section{Dependence of the Hamiltonian on vortex positions} \label{ap:inv}

In this appendix we will show that the GL free energy Hamiltonian in 
Eq.~(\ref{eq:ham})
depends only on the positions of the zeros in the order parameter on the 
sphere. First, from Eq.~(\ref{eq:prod}) we have
\begin{eqnarray}
\psi(\theta,\phi)&=&C\prod_{i=1}^N[\sin(\theta/2)
e^{i\phi}-z_i\cos(\theta/2)]\nonumber\\
&=&D\prod_{i=1}^N[a_i\sin(\theta/2)e^{i\phi}-b_i\cos(\theta/2)]\nonumber\\
&=&D\prod_{i=1}^Nf_i(\theta,\phi),
\end{eqnarray}
with $D=C\prod_{i=1}^N z_i/b_i$ and $b_i=a_iz_i$. If $f_i=0$ when 
$\theta=\theta_i$ and $\phi=\phi_i$ then we may choose:
\begin{eqnarray}
a_i&=&\cos(\theta_i/2)e^{-i\phi_i/2}\nonumber\\
b_i&=&\sin(\theta_i/2)e^{i\phi_i/2}.
\end{eqnarray}
  
Now, the GL-LLL free energy is a functional that depends only on the value of
$|\psi|^2=|D|^2|\prod_{i=1}^N|f_i|^2$ at each point on the sphere. If we can 
show that $|f_i(\theta,\phi)|^2$ depends only on the relative position of 
the $i$th zero then the Hamiltonian will be invariant with respect to 
coordinate rotations of all the zeros:
\begin{eqnarray}
|f_i(\theta,\phi)|^2&=&\biggl|
\cos(\theta_i/2)\sin(\theta/2)e^{i(\phi-\phi_i)/2}
\nonumber\\
&&\hspace{2cm}
-\sin(\theta_i/2)\cos(\theta/2)e^{-i(\phi-\phi_i)/2}\biggr|^2\nonumber\\
&=&\frac{1}{2}[1-\cos(\theta)\cos(\theta_i)-
\sin(\theta)\sin(\theta_i)\cos(\phi-\phi_i)]\nonumber\\
&=&\frac{1}{2}[1-\cos{\theta'(i)}]\nonumber\\
&=&\sin^2[\theta'(i)/2]
\end{eqnarray}
where $\theta'(i)$ is the polar angle measured from the axis through 
the $i$th zero to the point $(\theta,\phi)$, defined by an analogous 
expression to Eq.~\ref{eq:theta'}. This is independent of the 
coordinate frame and so we have shown that $|\psi(\theta,\phi)|^2$ 
depends only on the relative positions of the vortices plus the overall 
amplitude.

\section{Details for Density Correlations}     \label{ap:struc} 

To find the reciprocal space density correlator from thermal averages
of the LLL basis state coefficients (see Eq.~(\ref{eq:recip})) we need the
integrals
\begin{eqnarray}
I_{p,q,l}^m =
h_ph_q\int_0^{\pi}\int_0^{2\pi}d\theta && d\phi 
\, e^{i(q-p)\phi}
\sin{\theta}
\sin^{p+q}({\theta /2})\nonumber\\
&&\times\cos^{2N-p-q}({\theta /2})
\, Y_l^m(\theta,\phi)  ,
\end{eqnarray}
 which may be written:
\begin{eqnarray}
I_{p,q,l}^m =&&
\delta_{p-m,q}{(-1)}^mh_ph_{p-m} {\left[ \frac{(2l+1)(l+m)!}
{4\pi (l-m)!}\right]}^{1/2}\times\nonumber\\
&&\hspace{-1cm}
\times\frac{4\pi p!(N-p+m)!}{(N+m+1)!m!}
\,_3F_2\left[
\begin{array}{l}
 m-l,m+l+1,p+1
\\ m+1,N+m+2  \end{array}\right] ,
\end{eqnarray}
where $\,_3F_2[a,b,c;d,e]$ is a generalized hypergeometric series. To find the
infinite temperature limit of $C_l^m$ we consider the case $m=0$ and
calculate the averages $\langle v_p v_q^*v_r v_s^*\rangle$ and
$\langle v_p v_q^*\rangle$ exactly in the limit $\alpha_T\rightarrow +\infty$
(this involves only Gaussian integrals). Our result is
\begin{eqnarray}  
C_{l,\,\infty}&=&
\frac{(2l+1)}{(N+1)^2 }\sum_{p=0}^N
{\,_3F_2\left[
\begin{array}{l}
 -l,l+1,p+1
\\ 1,N+2  \end{array}\right]}^2 \nonumber\\
&=& \frac{(N!)^2}{(N-l)!(N+l+1)!}.\label{eq:infinite}
\end{eqnarray}
(We are indebted to T. Blum for this final result.)
This is consistent with the fact that 
in the limit $N\rightarrow \infty$, the infinite temperature
correlation function should be identical with that for an infinite 
plane,\cite{humac} and
the expression in Eq.~(\ref{eq:infinite}) should tend to $\exp{(-l^2/N)}$.
\end{multicols}

\widetext

\end{document}